\documentclass[fleqn,usenatbib]{mnras}
\usepackage{newtxtext,newtxmath}

\usepackage[T1]{fontenc}

\DeclareRobustCommand{\VAN}[3]{#2}
\let\VANthebibliography\thebibliography
\def\thebibliography{\DeclareRobustCommand{\VAN}[3]{##3}\VANthebibliography}

\usepackage{graphicx}	
\usepackage{amsmath}	

\usepackage{amssymb}	
\usepackage{color}      
\usepackage{array}
\usepackage[flushleft]{threeparttable}
\usepackage{booktabs,caption,soul}

\usepackage{scalerel,tikz}
\usetikzlibrary{svg.path}
\definecolor{orcidlogocol}{HTML}{A6CE39}
\tikzset{orcidlogo/.pic={
 \fill[orcidlogocol] svg{M256,128c0,70.7-57.3,128-128,128C57.3,256,0,198.7,0,128C0,57.3,57.3,0,128,0C198.7,0,256,57.3,256,128z};
 \fill[white] svg{M86.3,186.2H70.9V79.1h15.4v48.4V186.2z}
 svg{M108.9,79.1h41.6c39.6,0,57,28.3,57,53.6c0,27.5-21.5,53.6-56.8,53.6h-41.8V79.1z M124.3,172.4h24.5c34.9,0,42.9-26.5,42.9-39.7c0-21.5-13.7-39.7-43.7-39.7h-23.7V172.4z}
 svg{M88.7,56.8c0,5.5-4.5,10.1-10.1,10.1c-5.6,0-10.1-4.6-10.1-10.1c0-5.6,4.5-10.1,10.1-10.1C84.2,46.7,88.7,51.3,88.7,56.8z};
}}
\newcommand\orcidicon[1]{\href{https://orcid.org/#1}{\mbox{\scalerel*{
\begin{tikzpicture}[yscale=-1,transform shape]
\pic{orcidlogo};
\end{tikzpicture}
}{|}}}}

\newcommand{\Msun}{\,\mathrm{M}_\odot}

\newcommand{\Oabu}{12+\log(\mathrm{O/H})}
\newcommand{\re}{R_e}
\newcommand{\hii}{\mathrm{H~\textsc{ii}}}

\newcommand{\lcor}{l_\mathrm{corr}}
\newcommand{\aref}[1]{\hyperref[#1]{Appendix~\ref{#1}}}

\defcitealias{li2021}{L21}
\defcitealias{li2023}{L23}
\defcitealias{li2024}{L24}

\title[Metallicity correlations in AGN hosts]{Comparing metallicity correlations in nearby non-AGN and AGN-host galaxies}

\author[S.-L. Li et al.]{Song-Lin Li,$^{1,2}$\thanks{E-mail: songlin.li@anu.edu.au}
Zefeng Li$^{\orcidicon{0000-0001-7373-3115}}$, $^{1, 2, 3}$\thanks{E-mail: zefeng.li@durham.ac.uk}
Emily Wisnioski$^{\orcidicon{0000-0003-1657-7878}}$, $^{1, 2}$
Mark R. Krumholz$^{\orcidicon{0000-0003-3893-854X}}$, $^{1, 2}$
and
\newauthor Sebasti\'an F. S\'anchez$^{\orcidicon{0000-0001-6444-9307}}$\, $^{4, 5}$
\\
$^{1}$Research School of Astronomy \& Astrophysics, Australian National University, Canberra, ACT 2611, Australia\\
$^{2}$ARC Centre of Excellence for All-Sky Astrophysics in 3 Dimensions (ASTRO 3D), Canberra, ACT 2611, Australia\\
$^{3}$Centre for Extragalactic Astronomy, Department of Physics, Durham University, South Road, Durham DH1 3LE, UK\\
$^{4}$Universidad Nacional Aut\'onoma de M\'exico, Instituto de Astronom\'\i a, AP 106, Ensenada 22800, BC, M\'exico\\
$^{5}$Instituto de Astrof\'\i sica de Canarias, V\'\i a L\'actea s/n, 38205, La Laguna, Tenerife, Spain
}

\date{Accepted XXX. Received YYY; in original form ZZZ}

\pubyear{2024}

\begin{document}
\label{firstpage}
\pagerange{\pageref{firstpage}--\pageref{lastpage}}
\maketitle

\begin{abstract}
The gas-phase metallicity distribution within galaxies records critical information about galactic evolution. In this work we investigate how active galactic nuclei (AGN) influence this distribution by measuring the two-point correlation functions of gas-phase metallicity in 95 non-AGN and 37 AGN-host galaxies from the Calar Alto Legacy Integral Field spectroscopy Area integral field spectrographic survey. We measure metallicity using a novel Bayesian method that properly includes both stellar and AGN contributions to emission line fluxes and allows us to measure metallicities in both AGN-host and non-AGN galaxies in a single, consistent framework. We find that the two-point correlation functions of both AGN-host and non-AGN galaxies are well-fit by a simple injection-diffusion model, and that the correlation lengths $\lcor$ we derive for the non-AGN galaxies are reasonably consistent with those obtained in earlier work. The AGN-host galaxies generally have smaller $\lcor$ than non-AGN galaxies at fixed stellar mass, but similar $\lcor$ at fixed star formation rate (SFR), suggesting that the primary effect of hosting an AGN in this sample is a reduction in SFR at fixed stellar mass, and that this in turn suppresses the correlation length. Our findings further indicate that, while both SFR and stellar mass are positively correlated with metallicity correlation length $\lcor$, the former is more fundamental, implying that fluctuations in the metallicity distribution within galaxies are driven more by short-term responses to physical processes such as star formation that can change much faster than a Hubble time.
\end{abstract}




\section{Introduction}

Gas-phase metallicity, regulated by galactic inflows, outflows, star formation, and stellar nucleosynthesis, records key processes of galaxy formation and evolution (e.g. \citealt{sharda2021,sharda2024}). Thanks to large spectroscopic sky surveys such as Sloan Digital Sky Survey (SDSS), a number of empirical scaling relations for metallicity have been uncovered that can help us understand these processes. One of the most famous empirical relations is the mass-metallicity relation (MZR), a positive correlation between a galaxy's stellar mass and its gas-phase metallicity \citep[e.g.,][]{tremonti2004,kewley2008,andrews2013,perez2013}. Beyond this primary dependence, a number of authors have also identified secondary dependences on parameters other than stellar mass. For example, metallicity anti-correlates with star formation rate (SFR) at fixed stellar mass, as first demonstrated by \citet{lara2010} and \citet{mannucci2010}. The correlation between stellar mass, SFR, and metallicity is likely more fundamental than the MZR, and is therefore known as the fundamental metallicity relation (FZR); however, for a contrasting perspective see \citet{sanchez2013,sanchez2017,sanchez2019} and \citet{barrera2017}. A secondary dependence of metallicity on atomic or molecular hydrogen mass has also been explored \citep{bothwell2013,bothwell2016,brown2018}. These scaling relations are also found at high redshifts, showing a clear evolution through cosmic time \citep[e.g.,][]{erb2006,maier2014,zahid2014}, and have recently extended to lower stellar masses and to redshifts up to 10 by James Webb Space Telescope (JWST) observations \citep{lim2023,nakajima2023,he2024}. 

The prevalence of the integral field spectroscopy (IFS) surveys of nearby galaxies, such as MaNGA \citep[Mapping Nearby Galaxies at Apache Point Observatory,][]{bundy2015}, SAMI \citep[Sydney-AAO Multi-object Integral-field spectrograph,][]{bryant2015}, and CALIFA \citep[Calar Alto
Legacy Integral Field Area,][]{sanchez2012}, makes it possible to explore spatially-resolved metallicity scaling relations. For example, there is a local correlation between stellar mass surface density, star formation surface density, and metallicity, which has led some authors to suggest that the MZR and FZR are simply the cumulative effects of a more fundamental local scaling relation, an interpretation that emphasises the importance of local properties such as gas fraction in regulating the metallicity \citep[e.g.,][]{rosales2012,barrera2016,teklu2020,baker2023}. An azimuthally averaged metallicity gradient is another property that can be measured from IFS data \citep{sanchez2014}. \citet{belfiore2017} and \citet{poetrodjojo2018} find that the metallicity gradients steepen with the stellar mass in local star-forming (SF) galaxy populations. If we envision increasing stellar mass as an evolutionary sequence, this steepening requires metal redistribution by gas flows toward outskirts or metal dilution by pristine gas inflow \citep[e.g.,][]{mott2013,maiolino2019}.

In addition to mean metallicities and radial gradients, studies of IFS metallicity maps have begun to explore more complex statistical characterisations of 2D metallicity fields. For example, \citet{sanchezm2020} explore the metallicity variations between spiral-arm and interarm regions. However, perhaps the simplest non-trivial statistic one can compute from IFS data is the two-point correlation function, which reveals the interplay between metal production and metal mixing. \citet{krumholz2018} propose a theoretical model for this statistic based on competition between supernova production and metal diffusion, and several authors have measured it, or closely related statistics, in small samples of very nearby galaxies \citep[e.g.,][]{Kreckel2019, Kreckel2020, Metha2021, Metha2022}. By using the CALIFA and AMUSING++ IFS surveys \citep{lopez2020}, \citet[][hereafter \citetalias{li2021}]{li2021} and \citet[][hereafter \citetalias{li2023}]{li2023} successfully achieve larger sample sizes ($>100$ galaxies) and a significantly wider range of galaxy mass, SFR, and morphological type than the earlier studies. They found that \citeauthor{krumholz2018}'s model provides good fits to the observed two-point correlation functions across the full sample. The key fitting parameter derived from this analysis is the correlation length ($\lcor$) of the metallicity field. Observed $\lcor$ are typically $\sim$kpc, and are positively correlated with stellar mass ($M_*$), SFR, and galaxy effective radius ($\re$). Which of these is the most fundamental is not clear, as these three parameters are all correlated with each other as well.

Most studies of gas-phase metallicities to date have excluded galaxies where active galactic nuclei (AGN) makes significant contributions to the luminosity. This exclusion has been necessary because most metallicity calibrators have been developed for $\hii$ regions ionised by stellar sources, and are not reliable for gas ionised by non-stellar sources that typically produce much harder spectra \citep{kewley2019b}. There are a few exceptions where authors have attempted to derive the metallicity for AGN \citep[e.g.,][]{carvalho2020,dors2020,dors2021} using metallicity calibrators developed specifically for AGN-like ionising spectra. However, these calibrators are limited in pure AGN-dominant galaxies and not applicable to  galaxies with significant contributions from $\hii$ regions. Moreover, because different metallicity diagnostics have substantial systematic differences \citep[][and the references therein]{kewley2019b}, it is not straightforward to compare metallicities derived with these AGN-specific methods to those derived for non-AGN galaxies using different methods. When studying how the presence of an AGN influences galaxy metallicities, we require a more general approach that uses a uniform set of line diagnostics to disentangle the relative contribution from the SF-driven and AGN-driven ionisation in a self-consistent way.

The need for such a method was a primary motivation for the development of \textsc{NebulaBayes}, a Bayesian code that estimates the gas properties including metallicity by comparing a set of observed emission-line fluxes to photoionisation model grids \citep{thomas2018}. By adopting a mixture of $\hii$-region grids and narrow-line region (NLR) grids where the gas is ionised by nuclear activity, \textsc{NebulaBayes} is able to measure metallicity in galaxies ranging from purely SF to Seyferts where AGN is the dominant excitation source. \citet{thomas2019} use \textsc{NebulaBayes} to investigate the MZR for local Seyfert galaxies with SDSS single-fibre spectra. \citet[][hereafter \citetalias{li2024}]{li2024} extend this work to the FZR by applying \textsc{NebulaBayes} to the MaNGA IFS survey. They find that AGN hosts generally have higher metallicities than non-AGN galaxies at fixed stellar masses, but that this is primarily due to their lower SFR compared to non-AGN galaxies, i.e.~the difference is a natural result of the existence of FZR.

Our primary goal in this work is to combine the Bayesian AGN-SF decomposition method developed in \citetalias{li2024} to the study of full 2D metallicity distributions and their two-point correlations using the statistical techniques pioneered by \citetalias{li2021} and \citetalias{li2023}. Our motivation is to investigate how the presence of an AGN influences the distribution of metals \textit{within} galaxies, in addition to their overall mean metallicities. For example, one might anticipate that AGN could influence the metallicity fields by triggering galactic-scale gas outflows, or that AGN activity might be correlated with merging events that are expected to increase the correlation lengths \citep[e.g.,][]{harrison2014,fischer2015,king2015,koss2018,comerford2024}. We aim to determine whether any such influence is visible in the data.

Our plan for the rest of this paper is as follows. In \autoref{sec:methods}, we introduce our sample selection procedures, the implementation of \textsc{NebulaBayes}, and how to measure correlation lengths from metallicity fluctuation maps. We present our results and discuss the implication of the results in \autoref{sec:results} and \autoref{sec:discussion}, respectively. Finally, we summarise our findings in \autoref{sec:concluciton}. 

Throughout the work, we adopt a \cite{chabrier2003} initial mass function (IMF) and a flat WMAP7 cosmology: $H_0=70.4 \,\mathrm{km\; s^{-1} \,Mpc^{-1}}$, $\Omega_M=0.27$, and $\Omega_{\Lambda}=0.73$ \citep{komatsu2011}.

\section{Methods}
\label{sec:methods}

In this section, we start with a brief introduction to the CALIFA survey from which we draw our sample, and to our sample selection criteria, in \autoref{sec:sample}. We then introduce our implementation of \textsc{NebulaBayes} and describe how we use it to obtain metallicity maps in \autoref{sec:NB}. In \autoref{sec:corr_func} and \autoref{sec:lcorr}, we explain our procedures to calculate galactic two-point correlation functions and fit them by Markov chain Monte Carlo (MCMC) sampling to acquire the correlation lengths $\lcor$, respectively. Finally, in \autoref{sec:mass_sfr} we discuss ancillary data on galaxy masses and SFRs that we use in our analysis.

\subsection{Sample selection}
\label{sec:sample}

CALIFA is an IFU survey using the 3.5 m telescope at the Calar Alto Observatory to obtain three-dimensional spectra of nearby galaxies ($0.0005 < z < 0.05$) covering a wide range of stellar mass, morphology, and environment \citep{sanchez2012,walcher2014}. The observations are performed with the Potsdam Multi Aperture Spectrograph (PMAS) in the PPak mode \citep{roth2005,kelz2006}. In this work, we use the V500 grating that covers 3750 - 7500 \AA $~$ with a typical spectral resolution $R \sim 850$ around 5000 \AA $\;$($\sim 360$ km s$^{-1}$). The PMAS instrument covers at least two effective radii ($\re$) of each target. The data reduction pipeline before DR3 ({\tt v2.2}) recovers a typical spatial resolution $\sim 2 \farcs 50$, corresponding to 760 pc at $z\sim 0.015$ \citep{sanchez2016}. The latest extended data release (eDR3) contains 895 galaxies and develops a new reduction algorithm ({\tt v2.3}) that enhances the spatial resolution to $\sim 1 \farcs 50$ \citep{sanchez2023}. Our parent sample is from eDR3. However, we still use the {\tt v2.2} pipeline to acquire spatially-resolved properties to allow direct comparison with the work of \citetalias{li2021}, who used this pipeline, and who demonstrate that the spatial resolution $\sim 2 \farcs 50$ available from it is sufficient to obtain accurate measurements of $\lcor$. The three-dimensional spectral cubes are fed into the spectroscopy analysis package {\tt pyPipe3D} to analyze the stellar populations and fit emission lines \citep{lacerda2022}.

\citet{sanchez2023} provide classifications for galaxies in the CALIFA eDR3 sample based on the locations on Baldwin, Phillips,
\& Terlevich (BPT) diagram \citep{baldwin1981,kewley2001} applied to the emission-line flux ratios extracted from central $1 \farcs 50$ of each target galaxy. In this work, we only use three out of six groups provided. The galaxies in these three categories are required to have a signal-to-noise ratio (SNR) greater than 3 in the H$\alpha$ emission line and greater than unity in the H$\beta$, [N~\textsc{ii}]$~\lambda 6583$, and [O~\textsc{iii}]$~\lambda 5007$ lines. With this restriction, the three groups we use are

\begin{enumerate}
 \item star forming (BPT-SF): galaxies in this class are characterised by an equivalent width in the H$\alpha$ emission line, EW(H$\alpha$), above 3 \AA, and by line ratios $\log$([O~\textsc{iii}]$/\mathrm{H}\beta)$ and $\log$([N~\textsc{ii}]$/\mathrm{H}\alpha)$ that place them below the maximum starburst line proposed by \citet{kewley2001} on the BPT diagram.

 \item weak AGN (wAGN): galaxies in this category have line ratios above the maximum starburst line, and EW(H$\alpha$) from 3~\AA~to 6~\AA.

 \item strong AGN (sAGN): these are galaxies whose line ratios place them above the maximum starburst line and that have EW(H$\alpha$) greater than 6 \AA.
\end{enumerate}

After selecting galaxies in these three categories, we apply a series of additional cuts to the sample. We remove galaxies with axis ratios smaller than 0.4; we calculate the axis ratio as $b/a = \sqrt{1-e^2}$, where $e$ is the eccentricity provided in the property summary catalog for eDR3. This cut excludes galaxies that are close enough to edge-on to suffer from significant dust attenuation and mixing of light from regions with different physical conditions. 

In addition to removing galaxies from the sample based on axis ratios, within each galaxy we follow \citetalias{li2024} by also masking spaxels where EW(H$\alpha$) is below 3 \AA, or wherever the H$\alpha$ or H$\beta$ SNR is smaller than 3. The EW(H$\alpha$) cut is intended to remove possible contamination from post-asymptotic giant branch (pAGB) stars \citep{cid2011,belfiore2016,lacerda2018}, whereas the Balmer-line only SNR criterion is to avoid a possible sample bias to lower metallicity when applying SNR cuts independently to multiple metal lines \citep{salim2014,thomas2019}. \citetalias{li2024} demonstrate that applying a higher SNR cut to Balmer lines does not change the results. Finally, after applying these pixel-by-pixel masks, we remove from the sample any galaxies where we have fewer than 500 non-masked spaxels, a number that \citetalias{li2021} find is the minimum required for accurate measurements of two-point correlation functions. The final sample we use after applying these filters consists of 97 BPT-SF galaxies, 3 wAGN galaxies, and 35 sAGN galaxies. 

Finally, we note that by using the BPT diagram to identify galaxies as AGN hosts, we introduce a potential bias: we might be more likely to mis-classify AGN-host galaxies with higher SFRs as non-hosts because their bright stellar-driven emission masks the AGN signature. This is a potential concern since, as we show below, we find that there is a systematic difference in the SFRs of AGN-host and non-host galaxies. To mitigate this concern, in \aref{app:heii} we follow \citet{tozzi2023} by searching for signatures of He~\textsc{ii} emission as an additional AGN diagnostic, one less subject to this bias. In the appendix we show that, while this check does cause us to re-classify two of our BPT-SF galaxies as AGN hosts, this is too small an effect to materially alter our conclusions. For consistency with prior work, we therefore continue to use the classification based on the BPT diagram explained above for the main body of this paper.

\subsection{Application of \textsc{NebulaBayes}}
\label{sec:NB}

\begin{figure*}
    \resizebox{17cm}{!}{\includegraphics{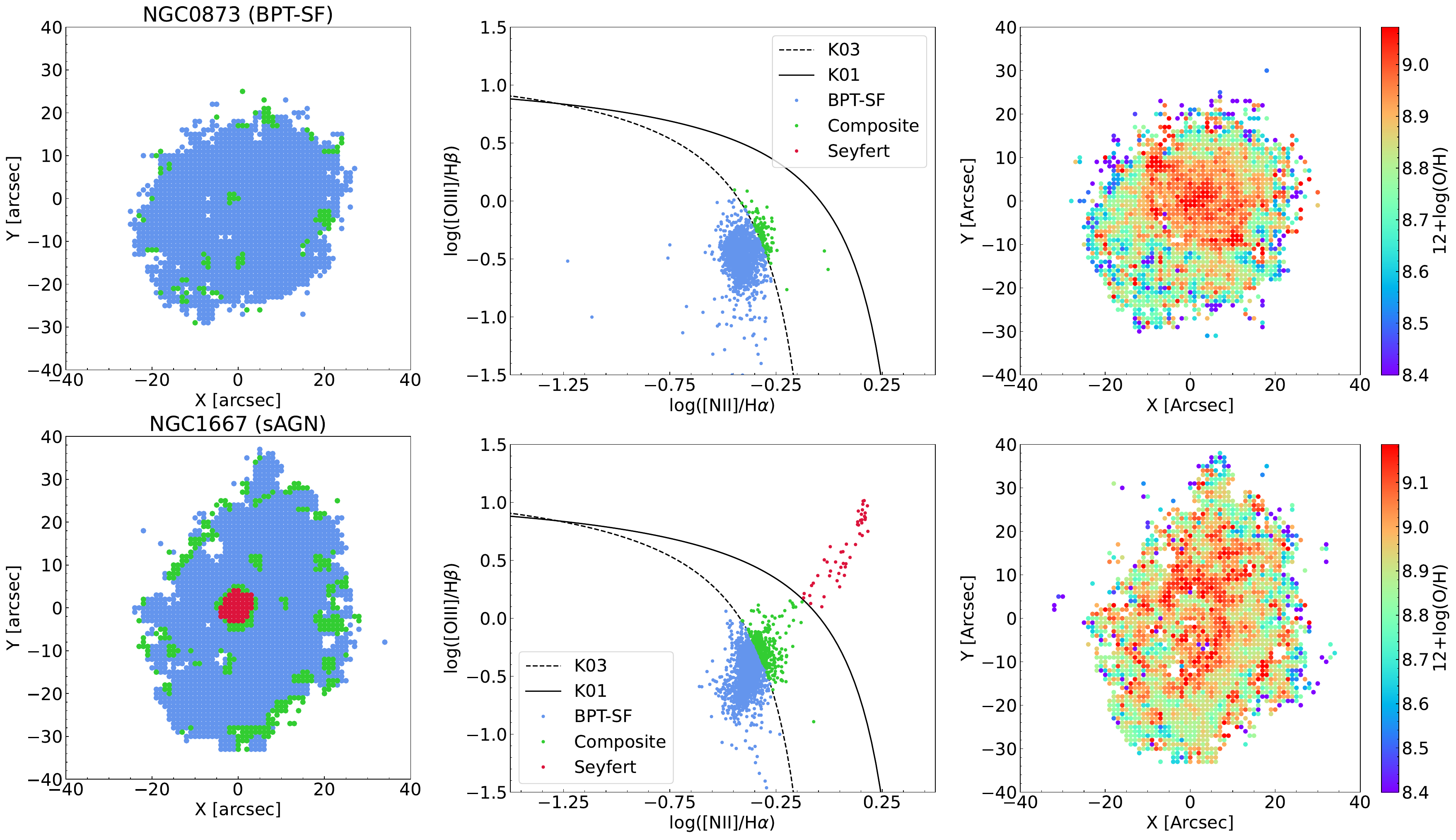}}
    \caption{A walkthrough of our analysis procedure for two sample galaxies: the upper row shows NGC0873, which is classified as a BPT-SF galaxy, while the lower row shows NGC1667, which is classified as sAGN. The left panels show the classifications we assign to each pixel; different colours correspond to different locations in the BPT diagrams shown in the middle panels, as indicated in the legend. The dashed lines in the middle panels are the demarcations proposed by \citet[][K03]{kauffmann2003} to distinguish pure SF galaxies from those with AGN contributions, while the solid lines show the theoretical maximum starburst locus proposed by \citet[][K01]{kewley2001}; pixels below the dashed lines are classified as BPT-SF, those above the solid lines are classified as Seyfert, and those between the two as Composite. The right panels show the final metallicity maps obtained from \textsc{NebulaBayes}.}
    \label{fig:bpt_example}
\end{figure*}

We use \textsc{NebulaBayes} \citep{thomas2018} to calculate metallicity maps for our sample. \textsc{NebulaBayes} estimates a series of Bayesian posterior parameters by comparing the observed emission-line fluxes with model grids produced by a photoionisation code such as \textsc{MAPPINGS} \citep{sutherland2017}. Instead of using the Solar-scaled scheme, we adopt the more realistic Galactic Concordance \citep{nicholls2017} scheme to set the relative abundance of metal elements as a function of gas-phase metallicity. In paticular, the nitrogen-to-oxygen ratio as a function of $\Oabu$, which determines the [N~\textsc{ii}]/[O~\textsc{ii}] ratio we use as a prior to estimate the metallicity in NebulaBayes, is obtained by fitting  nearby stars and nebulae with a combination of two linear functions that address both the primary and secondary nitrogen. In this work, we use the same procedure and set of photoionisation model grids as in \citetalias{li2024}. Here we briefly summarise the procedures and refer readers to \citetalias{li2024} for more details.

To apply \textsc{NebulaBayes}, we first classify spaxels as BPT-SF, Composite, or Seyfert based on their position in the BPT diagram; this classification is an efficiency measure, since it lets us avoid carrying out computationally-expensive fits using arbitrary combinations of stellar and AGN ionisation models in spaxels where there is clearly no AGN contribution. Note that these spaxel-by-spaxel classifications are not identical to the overall galaxy classifications discussed in \autoref{sec:sample}, which are based on integrated central spectra, so that a galaxy that is classified as BPT-SF may still contain spaxels with an AGN contribution, and vice versa. \autoref{fig:bpt_example} shows the spaxel-by-spaxel classifications for one example galaxy classified as BPT-SF and one as sAGN. The left panels show the spaxel classifications, while the middle panels show the locations of each spaxel in the BPT diagram. Spaxels that fall below the demarcation line proposed by \citet[dashed lines]{kauffmann2003} are classified as BPT-SF, and are assumed to be ionised by radiation from massive stars. Spaxels above the theoretical maximum starburst line proposed by \citet[solid lines]{kewley2001} are classified as Seyfert, and are presumably dominated by AGN ionisation. Spaxels lying between the two lines are classified as Composite, and likely mix both sources of ionisation.

For BPT-SF spaxels, we assume emission is solely from $\hii$ regions, and we therefore fit to $\hii$-region-only models, which have three free parameters: the oxygen abundance $\Oabu$ (i.e. the gas-phase metallicity), the ionisation parameter at the inner edge of the modeled nebula $\log U_\hii$, and the gas pressure $\log(P/k)$. We use the `Line-ratio prior mode' in \textsc{NebulaBayes} for these spaxels; in this mode, we run \textsc{NebulaBayes} three times, with each run using only one emission-line ratio that is sensitive to one parameter: $\mathrm{[N~\textsc{ii}]}~\lambda 6583/\mathrm{[O~\textsc{ii}]}~\lambda \lambda 3626, 29$ for $\Oabu$ \citep{kewley2002}, $\mathrm{[O~\textsc{iii}]}~\lambda 5007/\mathrm{[O~\textsc{ii}]}~\lambda \lambda 3626, 29$ for $\log U_\hii$ \citep{kewley2002,kobulnicky2004}, and $\mathrm{[S~\textsc{ii}]}~\lambda 6716/\mathrm{[S~\textsc{ii}]}~\lambda 6731$ for $\log(P/k)$ \citep{kewley2019a}. We correct dust reddening by using the difference between the intrinsic H$\alpha$/H$\beta$ ratio of each point of the model grid and the observed H$\alpha$/H$\beta$ ratio to deredden all lines before comparing the data to that model (see \citetalias{li2024} for details). This yields three sets of posterior likelihoods for each point of our model grid, which we then multiply together to get the final posterior, with the [N~\textsc{ii}]/[O~\textsc{ii}] weighted three times as much as other line ratios since metallicity is the key parameter of interest in this work. As discussed in \citetalias{li2024}, this simplified procedure for BPT-SF spaxels yields results for the metallicities that are very close to those obtained by carrying out a full run of \textsc{NebulaBayes} using all available emission lines, but at a small fraction of the computational cost.

For Composite and Seyfert spaxels, we fit a full model whereby we assume that the emission in that spaxel represents a linear combination of an $\hii$-region component and a narrow-line region (NLR) component ionised by a theoretical AGN spectrum generated using the \textsc{OXAF} code \citep{thomas2016}. We adopt a power-law index $\Gamma=-2.0$ for the non-thermal component of the AGN spectrum and assume that this component contributes a fraction $p_\mathrm{NT}=0.15$ of the total AGN luminosity. We fix the ionisation parameter of $\hii$ regions in these spaxels to $\log U_\hii=-3.25$, since this is the median $\log U_\hii$ of BPT-SF spaxels found in \citetalias{li2023}, and we assume that each spaxel is characterised by a single oxygen abundance and gas pressure. Under these assumptions we have five free parameters: $\log(P/k)$, $\Oabu$, the ionisation parameter of the NLR $\log U_\mathrm{NLR}$, the energy at which the Big Blue Bump component from the AGN accretion disc has its peak $\log (E_\mathrm{peak}/\mathrm{keV})$, and the fractional contribution of $\hii$ regions to the H$\alpha$ luminosity $f_\hii$. In addition, since all spaxels within the same galaxy should see the same AGN spectrum, we require $\log (E_\mathrm{peak}/\mathrm{keV})$ to be the same for all spaxlels in a single galaxy; we adopt a value of $-1.35$ for this parameter for spaxels in BPT-SF galaxies \citep{thomas2019}, while for wAGN and sAGN galaxies we choose a value by selecting Seyfert spaxels within central $3''$ and running \textsc{NebulaBayes} with $\log (E_\mathrm{peak}/\mathrm{keV})$ left as a free parameter, and fix the value of $\log (E_\mathrm{peak}/\mathrm{keV})$ for that galaxy to the modal outcome over these spaxels. We then run \textsc{NebulaBayes} three times to determine the remaining four parameters. The first two runs are in `Line-ratio prior mode' with [N~\textsc{ii}]/[O~\textsc{ii}] and $\mathrm{[S~\textsc{ii}]}~\lambda 6716 / \mathrm{[S~\textsc{ii}]}~\lambda 6731$, while the final run uses the `full-line likelihood mode' on the 10 emission lines: $\mathrm{[O~\textsc{ii}]}~\lambda \lambda3726, 29$, $\mathrm{[Ne~\textsc{iii}]}~\lambda 3869$, $\mathrm{H}\beta$, $\mathrm{[O~\textsc{iii}]}~\lambda 5007$, $\mathrm{He~\textsc{i}}~\lambda 5876$, $\mathrm{[O~\textsc{i}]}~\lambda 6300$, $\mathrm{H}\alpha$, $\mathrm{[N~\textsc{ii}]}~\lambda 6583$, $\mathrm{[S~\textsc{ii}]}~\lambda6716$, and $\mathrm{[S~\textsc{ii}]}~\lambda 6731$. We obtain the final posterior by multiplying the posteriors produced by these three runs together. Dust reddening is handled exactly as for BPT-SF spaxels.

The right panels of \autoref{fig:bpt_example} show the metallicity maps we derive by this procedure for our two example galaxies. Both the two galaxies show a clear negative metallicity gradient. We apply this procedure to the full sample to generate a metallicity map for each galaxy.

\subsection{Measuring the correlation functions}
\label{sec:corr_func}

\begin{figure*}
    \resizebox{17cm}{!}{\includegraphics{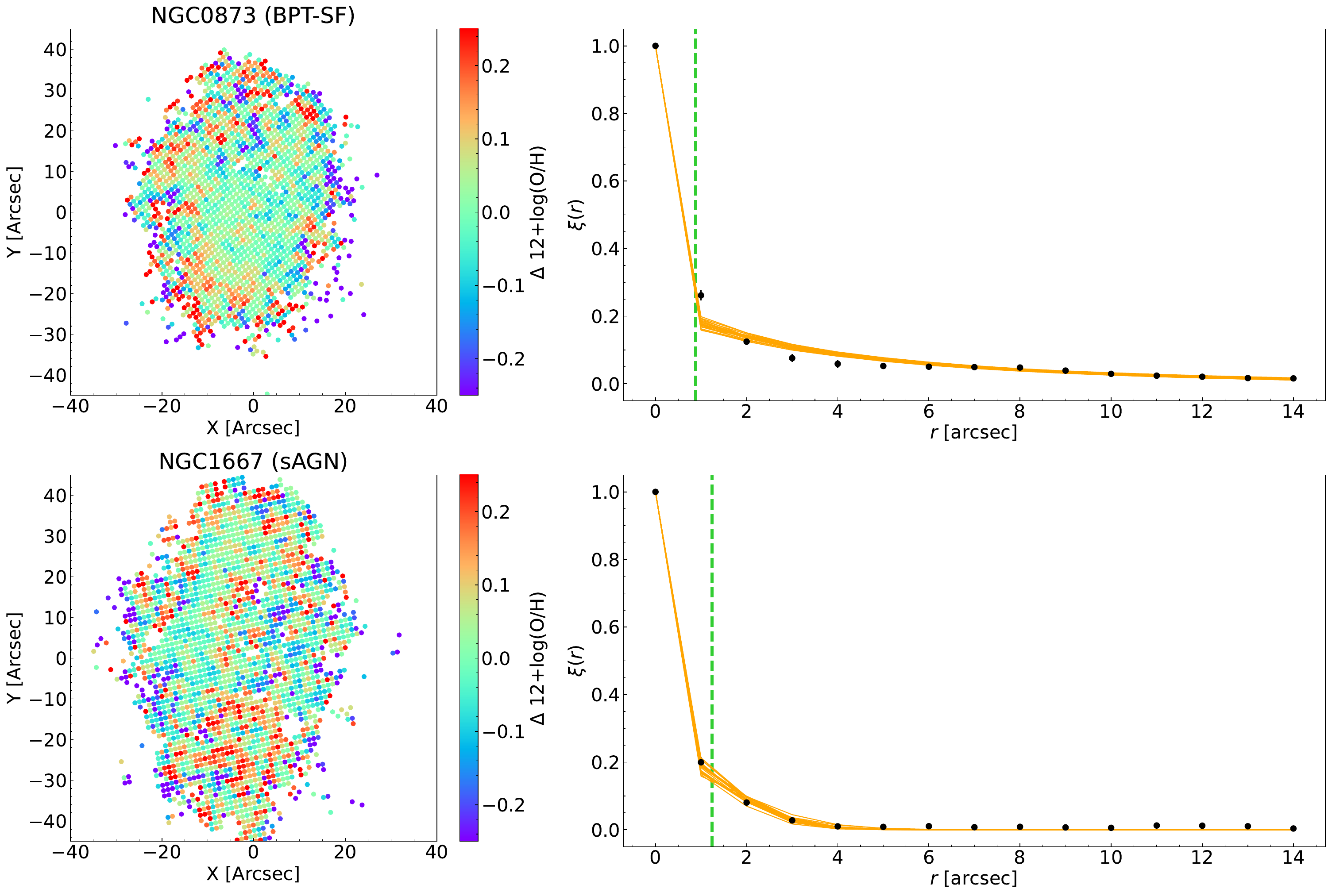}}
    \caption{Left: the metallicity fluctuation maps of the same two example galaxies shown in \autoref{fig:bpt_example}. Right: the black dots and the error bars (which are too small to be seen for most points) show the observed two-point correlation functions and the corresponding uncertainties, comparing with 20 MCMC-fit models (orange lines) randomly selected from the converged sampling chains. The green dashed lines indicate the the median values of the posterior PDF of $\sigma_{\rm beam}$. See \autoref{sec:lcorr} for details.}
    \label{fig:corr_example}
\end{figure*}

We next measure the two-point correlation functions for our metallicity maps following the procedure outlined by \citetalias{li2021}; as with \autoref{sec:NB}, we only summarise here, and refer readers to the original paper for full details of the procedure. Our first step is to generate metallicity fluctuation maps from the metallcity maps output by \textsc{NebulaBayes}. To do so we deproject and rotate the metallicity maps to acquire a circular profile with the major axis of the original map along the $x$ axis. The required transformation from the original coordinate $(x, y)$ of each spaxel to the new one $(x', y')$ is given by
\begin{equation}
\left[\begin{array}{c}
x'\\
y'
\end{array}\right]
=
\left[\begin{array}{cc}
\cos\theta & \sin\theta \\
-\sin\theta/\cos i & \cos\theta/\cos i
\end{array}\right]
\left[\begin{array}{c}
x\\
y
\end{array}\right],
\end{equation}
where $\theta$ is the position angle adopted from the eDR3 catalog, $i$ is the inclination angle, determined by
\begin{equation}
\cos^2 i = \frac{(b/a)^2-q_0^2}{1-q_0^2},
\label{equ:cosi}
\end{equation}
where the $b/a$ is the axis ratio, and $q_0=0.13$. \citetalias{li2023} show that adopting a larger $q_0$ does not change the results. 

We then calculate the metallicity fluctuation of the $i$th spaxel $Z'_i$ by subtracting the mean metallicity $\overline{Z_r}$, calculated within an annular bin with the same distance $r$ to the galactic center as the $i$th spaxel and with a bin width of $1''$, which is the spaxel size of CALIFA datacubes. The left panels of \autoref{fig:corr_example} show the rotated and deprojected metallicity fluctuation maps of the same examples from \autoref{fig:bpt_example}.

The two-point correlation function of the metallicity fluctuation map is
\begin{equation}
    \xi(\mathbf{r}) = \frac{\left\langle Z'(\mathbf{r + r'}) Z'(\mathbf{r}') \right\rangle}{\left\langle Z'(\mathbf{r}')^2\right\rangle},
\end{equation}
where $Z'(\mathbf{x})$ is the metallicity fluctuation we obtained above at position $\mathbf{x}$ in the map, and the angle brackets $\left\langle\cdot\right\rangle$ denote averaging over the dummy position variable $\mathbf{r}'$. In practice, we integrate out the direction of the vector $\mathbf{r}$ since we are only interested in the scalar separation $r$ between two spaxels. We compute the two-point correlation function averaged over bins of separation $r$ that are $1''$ wide, equal to the spaxel size of CALIFA datacubes. We compute the two-point correlation function of the $n$th bin as
\begin{equation}
    \xi_n = \frac{\sigma_{Z'}^{-2}}{N_n} \sum_{r_n < r_{ij} \leq r_{n+1}} Z'_i Z'_j,
\end{equation}
where $r_n = n\times(1'')$, the sum runs over the $N_n$ spaxel pairs $(i, j)$ whose separations lie in the range $r_n < r_{ij} \leq r_{n+1}$, and 
\begin{equation}
    \sigma_{Z'}^2 = \frac{1}{N_p} \sum_{i=1}^{N_p} {Z'_i}^2
\end{equation}
is the sum of the variances of all $N_p$ spaxels in the fluctuation map. We evaluate the two-point correlation to a maximum separation of $15''$, corresponding to 4 kpc at $z \sim 0.015$, the mean redshift of the CALIFA sample.

To estimate the uncertainty of the two-point correlation function, we randomly draw a value for the flux of each emission line in each pixel of the observed map from a Gaussian distribution with a mean and standard deviation set equal to the central estimate and estimated error for that spaxel in the observed map. This yields a new flux map. We mask spaxels in this map for which the flux is negative in the H$\alpha$, H$\beta$, [O~\textsc{iii}], [N~\textsc{ii}], [O~\textsc{ii}], $\mathrm{[S~\textsc{ii}]}~\lambda 6716$, or $\mathrm{[S~\textsc{ii}]}~\lambda 6731$ lines\footnote{For other weak lines with negative flux used for the Composite and Seyfert spaxels, we assign a flux of 5 per cent of the error. This assignment has little weight in our likelihood calculation, but satisfies the formal requirement that all fluxes input to \textsc{NebulaBayes} be positive.}. We then run \textsc{NebulaBayes} on the resulting synthetic map of line fluxes exactly as described in \autoref{sec:NB} to generate a synthetic metallicity map, and compute the two-point correlation of this metallicity map as described above. We repeat this procedure 20 times, yielding 20 two-point correlation functions. We take the standard deviation of these 20 realizations at each separation bin as our estimate of the uncertainty for that bin.

The black dots in the right panels of \autoref{fig:corr_example} show the two-point correlation functions of the same example galaxies used in \autoref{fig:bpt_example}. Note that $\xi(0)$ is always unity, since every spaxel is perfectly correlated with itself. Error bars are shown, but for most points they are invisibly small because each bin represents a sum over a large number of spaxel pairs, and this summation averages down the final uncertainties.

\subsection{Estimation of the correlation length}
\label{sec:lcorr}

\begin{figure}
{\includegraphics[width=\columnwidth]{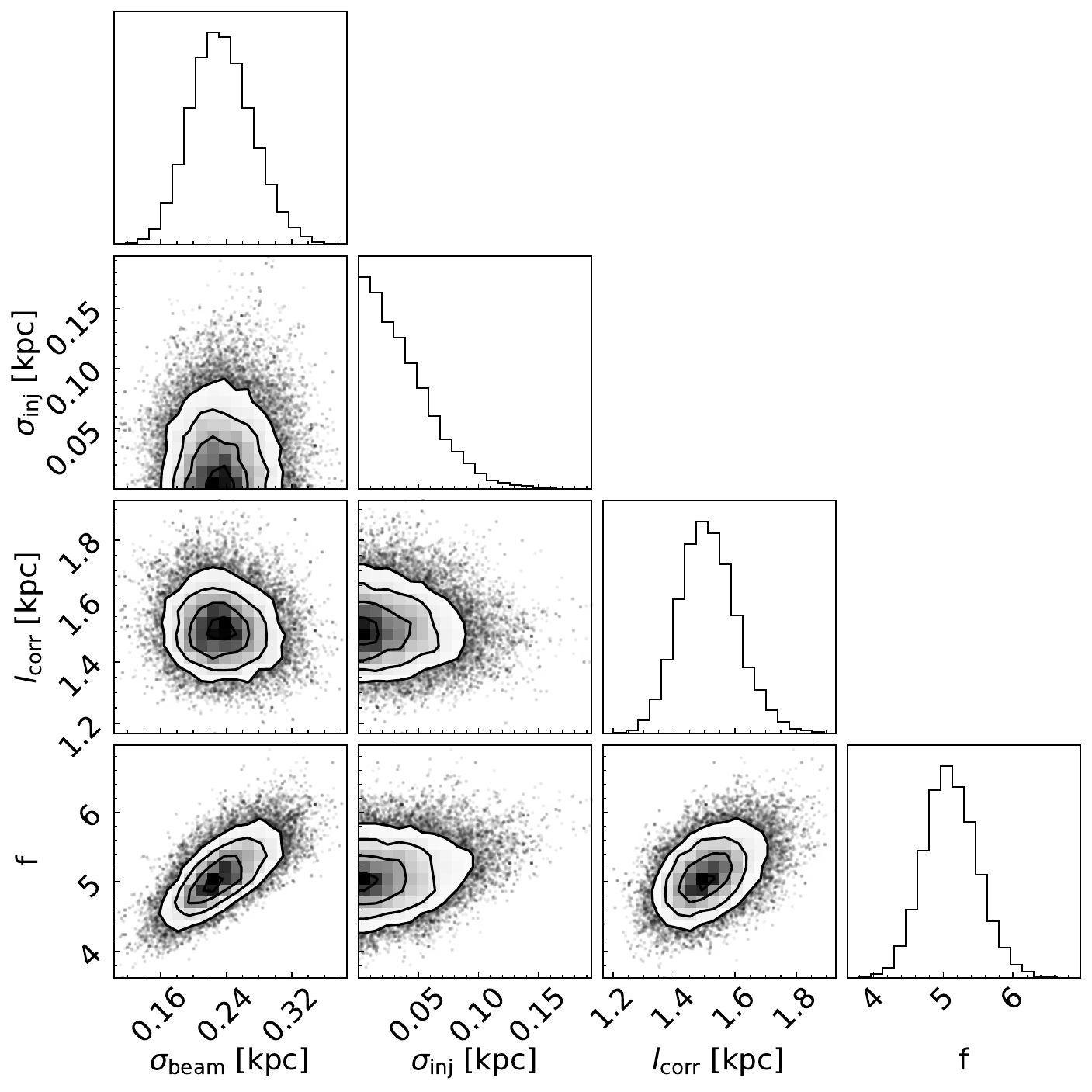}}
    \caption{The posterior distribution of four parameters describing the two-point correlation function for NGC0873, a BPT-SF example. The corresponding BPT map, BPT diagram, metallicity map, metallicity fluctuation map, and two-point correlation function are shown in \autoref{fig:bpt_example} and \autoref{fig:corr_example}. The contours represent the 0.5, 1, 1.5, and 2$\sigma$ confidence levels, i.e., the 1$\sigma$ contour encloses 39.3\% of the total probability, etc.}
    \label{fig:tri_example}
\end{figure}

\citetalias{li2021} and \citetalias{li2023} find that the two-point correlation functions of non-AGN host galaxies in the CALIFA and AMUSING++ samples are well-fit by the injection-diffusion model proposed by \citet{krumholz2018} with additional terms to account for the effects of beam smearing and observational noise. The functional form for this model is (see Appendix D of \citetalias{li2021} for details)
\begin{eqnarray}
\lefteqn{\xi_{\rm model}(r) = \frac{2}{\ln\left(1 + \frac{2\lcor^2}{\sigma_0^2/2}\right)
} \left[\frac{\Theta(r-\ell_{\rm pix})}{f} + \Theta(\ell_{\rm pix}-r)\right]
}
\nonumber \\
& &
\int_0^\infty e^{-\sigma_0^2 a^2/2} \left(1 - e^{-2 \lcor^2 a^2}\right) \frac{J_0(ar)}{a} \, da,
\end{eqnarray}
where $\sigma_0^2 = \sigma_{\rm beam}^2 + 2 \sigma_{\rm inj}^2$. $\sigma_{\rm beam}$ is the observational beam size, $\sigma_{\rm inj}$ is the width over which supernovae inject metals, $\lcor$ is the correlation length, $f$ is the ratio of the variance of the metallicity fluctuation map including observational uncertainties to the real variance, $\ell_\mathrm{pix}$ is the pixel scale of the fluctuation map, $J_0(x)$ is the modified Bessel function of the first kind of order 0, and $\Theta(x)$ is the Heaviside step function. The terms involving $\Theta(x)$ account for the fact that the errors within each spaxel of the map are perfectly correlated with themselves, but are totally uncorrelated with the errors in different spaxels.

The free parameters in this model are $\sigma_{\rm beam}$, $\sigma_{\rm inj}$, $\lcor$, and $f$, and we fit these to our measured correlation functions using MCMC sampling with the python package \textsc{emcee} \citep{foreman2013}. We set several priors for these four parameters. All parameters are required to be positive, so we set priors for non-positive values to zero. For $\sigma_{\rm beam}$, we cross-match our sample with the point spread function (PSF) catalogue from \citet{sanchez2016}, in which they measure the PSF by comparing the reconstructed $g$-band images from CALIFA datacubes to SDSS images; this catalogue provides beam sizes for 96 of the 135 galaxies in our sample. Because we deproject the metallicity maps when generating the metallicity fluctuation maps in \autoref{sec:corr_func}, the beam sizes in our science maps are stretched by a factor of $1/\cos i$ along the galactic minor axis, where $i$ is the inclination angle calculated from \autoref{equ:cosi}. For the 96 galaxies with PSF measurements $\sigma_\mathrm{PSF}$ in the \citeauthor{sanchez2016} catalogue, we assign a Gaussian prior with a mean of $\sigma_\mathrm{PSF}(1+\cos i)/(2\cos i)$, the arithmetic mean of the major and minor axis PSF sizes, and a standard deviation of $\sigma_\mathrm{PSF}(1-\cos i)/(2\cos i)$, which is half of the difference between the major and minor axis PSF sizes. This prior covers the plausible range of a circular beam size due to the beam-stretching because of the deprojection. For galaxies without a PSF measurement, we assign a Gaussian prior with a mean of $1\farcs 06 \times (1+\cos i)/(2\cos i)$ and a standard deviation of $0\farcs 14 \times (1+\cos i)/(2\cos i)$, where $1\farcs 06$ and $0\farcs 14$ are the median and the 1$\sigma$ scatter of the distribution of the CALIFA beam sizes. In \aref{app:sigma_beam} we test the sensitivity of our results to this treatment of beam smearing by testing the alternative approaches of fixing $\sigma_{\rm beam}$ to either $\sigma_\mathrm{PSF}$ or $\sigma_\mathrm{PSF}/\cos i$, i.e., the the largest and smallest plausible values. We find that doing so does not alter the qualitative results.

For $f$, we assign a flat distribution in logarithmic space up to a maximum $f=40$. We set this upper limit to be large enough so that for values of $f$ near the upper limit any real metallicity fluctuations are completely hidden in the noise and therefore cannot be estimated. A value of $f$ near the upper limit effectively counts as a non-detection of any correlation in the metallicity fluctuation map. Finally, we adopt a flat prior on $\lcor$ from zero to a physical distance corresponding to an angular size of $80''$, the same as the bundle size of CALIFA. Values of $\lcor$ larger than this are unmeasurable given CALIFA's field of view.

Following \citetalias{li2023}, for each galaxy we use 100 walkers in the MCMC and run for 1000 steps. We treat the first 500 steps as the burn-in period for the walkers to converge, and only keep the last 500 steps. The posterior PDF of each parameter thus consists of 50,000 sampling points. \autoref{fig:tri_example} shows the posterior distribution of four parameters describing the two-point correlation function for the BPT-SF example, NGC0873, shown in \autoref{fig:bpt_example} and \autoref{fig:corr_example}. The values of $\sigma_{\rm beam}$, $\lcor$, and $f$ are well constrained, but we are only able to obtain an upper limit on $\sigma_{\rm inj}$. As discussed in \citetalias{li2021}, this is expected, since $\sigma_{\rm inj}$ is expected to be $<100$ pc, always smaller than the CALIFA beam size. Given that $\sigma_{\rm beam}$ and $\sigma_{\rm inj}$ are combined into $\sigma_0$, at our spatial resolution we only expect to be able to obtain an upper limit on $\sigma_{\rm inj}$. In the right panels of \autoref{fig:corr_example} we randomly select 20 MCMC-fit models from the converged chains and plot them on top of the observed two-point correlation functions (orange lines). They fit the observed data quite well for both the BPT-SF and AGN-host galaxies. We also show the median values of the posterior PDF of $\sigma_{\rm beam}$ as green dashed lines for reference. In what follows, we take the median value of the posterior PDF of the $\lcor$ as our central estimate for each galaxy and the half of the 16th to 84th percentile range as the corresponding uncertainty. 

While this procedure yields well-behaved and narrowly-peaked posterior PDFs for 132 of our sample of 135 galaxies, we find that three galaxies show long tails in their posterior distributions of $\lcor$ and $f$. Examination of the two-point correlation functions for these galaxies shows that they are close to 0, with large errors, except for the bin at $\xi(0)$ that is always unity. We discuss these galaxies further in \aref{app:fail}, where we show that our analysis fails for these galaxies because we measure essentially no meaningful correlation, and our fitting procedure cannot distinguish between two possible explanations: either these galaxies have $\lcor$ significantly smaller than the beam size or observational errors so large that any real correlations are unmeasurable against the noise. Given the degeneracy of these two possibilities, and the fact that it affects only 2\% of our sample, we simply remove these three galaxies from our analysis, leaving a final sample consisting of 95 BPT-SF galaxies, 3 wAGN galaxies, and 34 sAGN galaxies with well-measured correlation lengths. Because we have so few wAGN in our sample, we merge the wAGN and the sAGN categories into a single AGN-host category in the following analysis.

\subsection{Ancillary data: stellar mass and SFR}
\label{sec:mass_sfr}

\begin{figure}
{\includegraphics[width=\columnwidth]{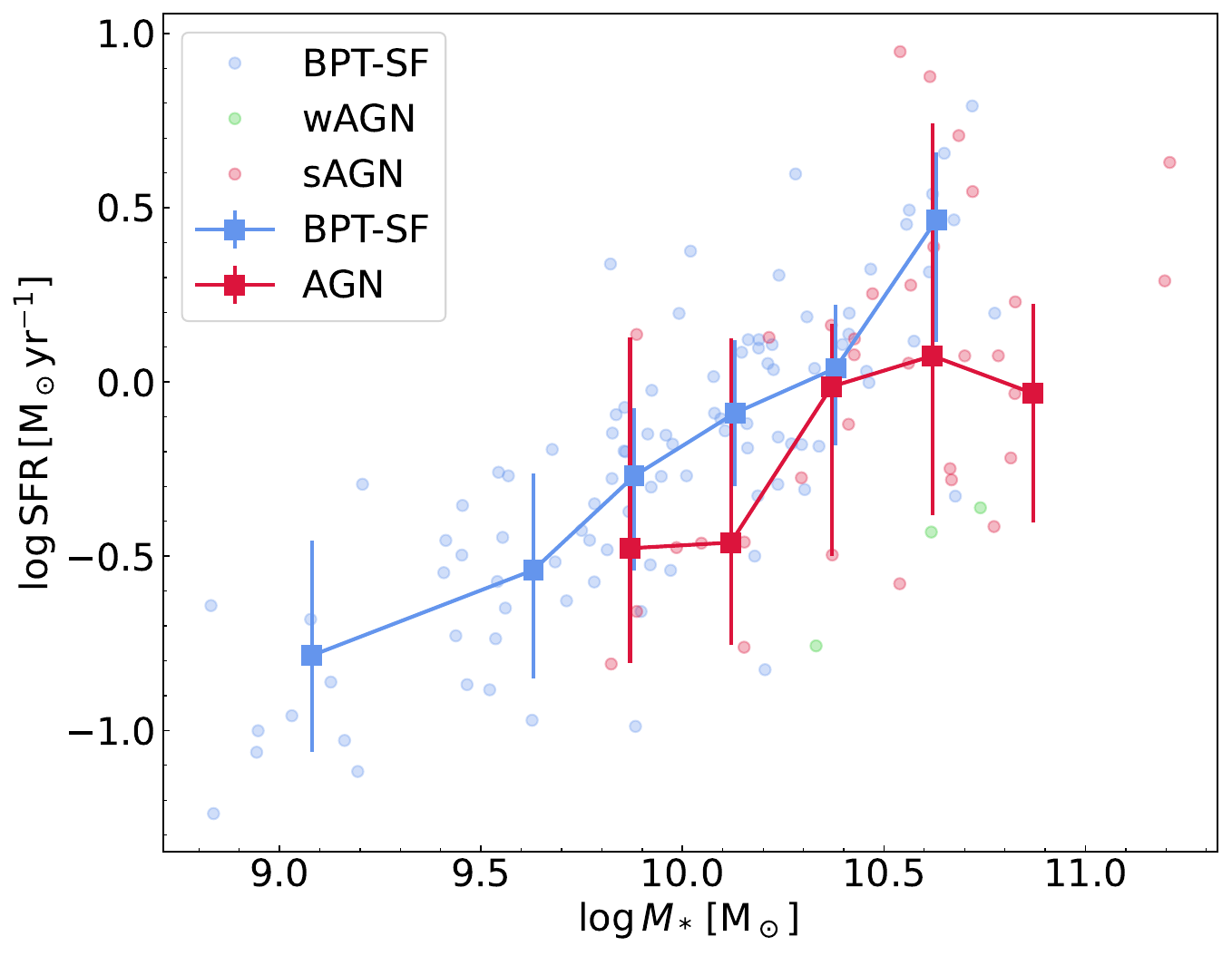}}
    \caption{SFR versus stellar mass for BPT-SF galaxies (blue dots), wAGN galaxies (green dots), and sAGN galaxies (red dots). The squares are the median SFR in each stellar mass bin for BPT-SF galaxies (blue), and AGN-host galaxies (red, combining the wAGN and sAGN groups). The error bars show the 1$\sigma$ scatter estimated via the procedure described in \autoref{sec:mass_sfr}. The stellar mass bins are the same for both BPT-SF and AGN-host galaxies, but the points have been offset slightly for readability.}
    \label{fig:sfms}
\end{figure}

In addition to the correlation length, we require ancillary data on other galaxy properties, in particular the stellar mass and star formation rate. We take these directly from the eDR3 catalogue. The stellar mass is estimated by fitting the integrated spectra with simple stellar population (SSP) templates, and the SFR is converted from the dust-corrected integrated H$\alpha$ luminosity \citep{sanchez2023}. The stellar mass and SFR in the eDR3 catalogue are derived assuming a \citeauthor{salpeter1955} IMF, and we correct both of them by 0.25 dex to the \citeauthor{chabrier2003} IMF that we use throughout this work \citep{bernardi2010}.

\citet{sanchez2023} point out that the conversion from the H$\alpha$ luminosity to SFR in the catalogue does not account for potential contributions to the ionising luminosity from sources other than massive stars, and thus might overestimate the true SFRs in galaxies with significant AGN contributions. However, we show in \aref{app:sfr_corr} that, while AGN contributions to individual pixels can be significant, contributions to the integrated ionising luminosity of the full galaxy, and thus potential errors in the inferred SFR, are small. Consequently, making a correction to the SFR for a potential AGN contribution does not produce qualitative any difference to our results. We therefore omit such corrections for simplicity.

\autoref{fig:sfms} shows SFR versus stellar mass for both the BPT-SF (blue) and AGN-host galaxies (red, combining wAGN and sAGN groups) in our sample. The squares show the median SFRs in each stellar mass bin. To estimate the 1$\sigma$ levels of the scatter, for each galaxy lying in one stellar mass bin, we randomly draw 1000 realizations from a Gaussian distribution with a standard deviation equal to the stated uncertainty on the SFR from the eDR3 catalogue. We then take the 84th and the 16th percentiles over all realisations for all galaxies lying in this stellar mass bin as the upper and the lower uncertainties, respectively. It is clear that AGN-host galaxies generally have lower SFR than BPT-SF (non-AGN) galaxies at fixed stellar mass, consistent with earlier studies of optically selected AGN samples \citep[e.g.,][]{schawinski2014,ellison2016,lacerda2020}.

\section{Results}
\label{sec:results}

Here we first carry out two tests of the robustness of our results against possible errors and biases, and then present the main findings of this study.

\subsection{Robustness tests}

\begin{figure}
{\includegraphics[width=\columnwidth]{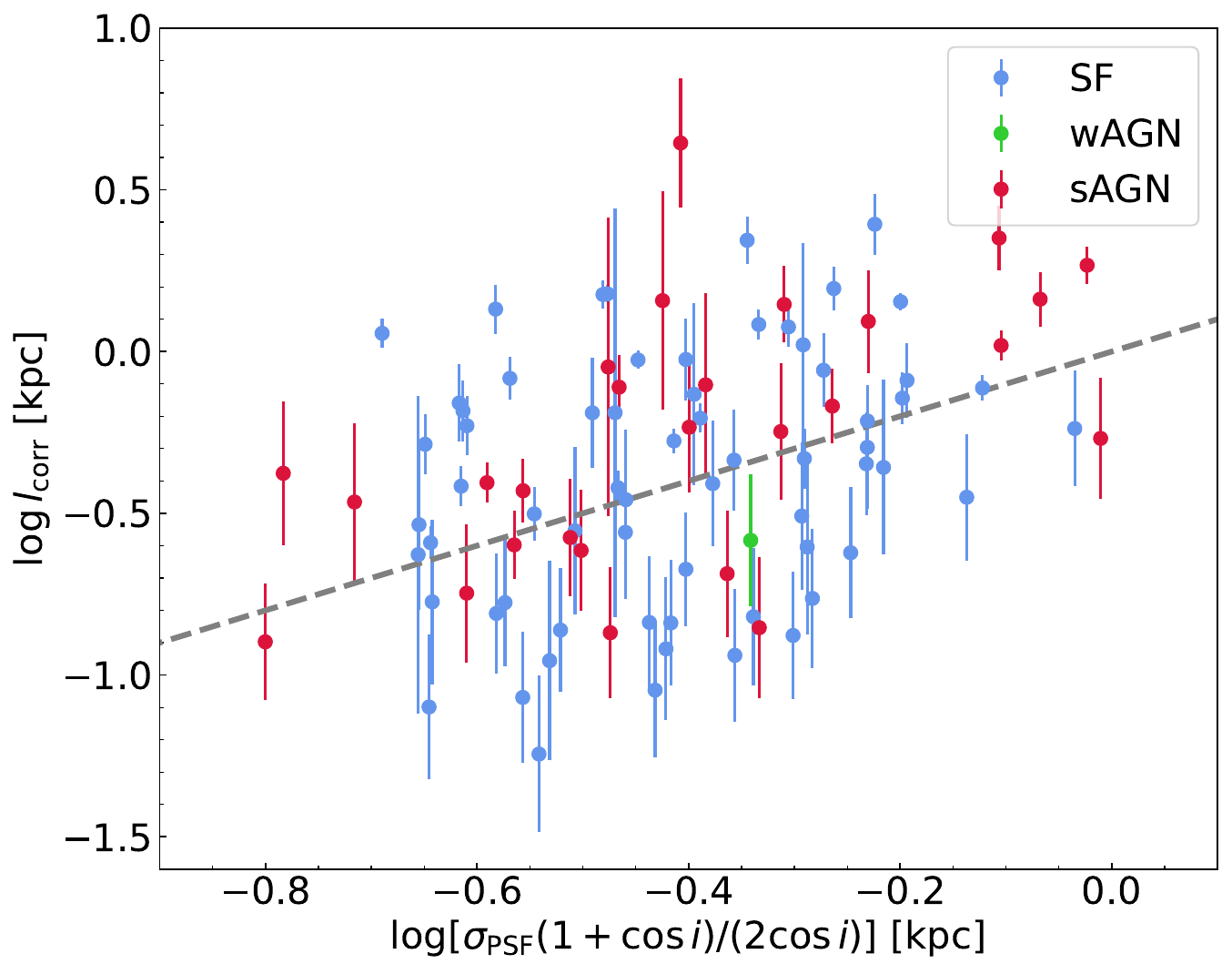}}
    \caption{The relation between estimated correlation lengths and the beam size $\sigma_\mathrm{PSF} (1+\cos i)/(2\cos i)$ for 96 galaxies with PSF measurements in CALIFA, where $\sigma_\mathrm{PSF}$ is the beam size in physical scale (kpc), and $i$ is the inclination angle calculated by \autoref{equ:cosi}. The dashed grey line shows the one-to-one relation.}
    \label{fig:beam}
\end{figure}

\begin{figure}
{\includegraphics[width=\columnwidth]{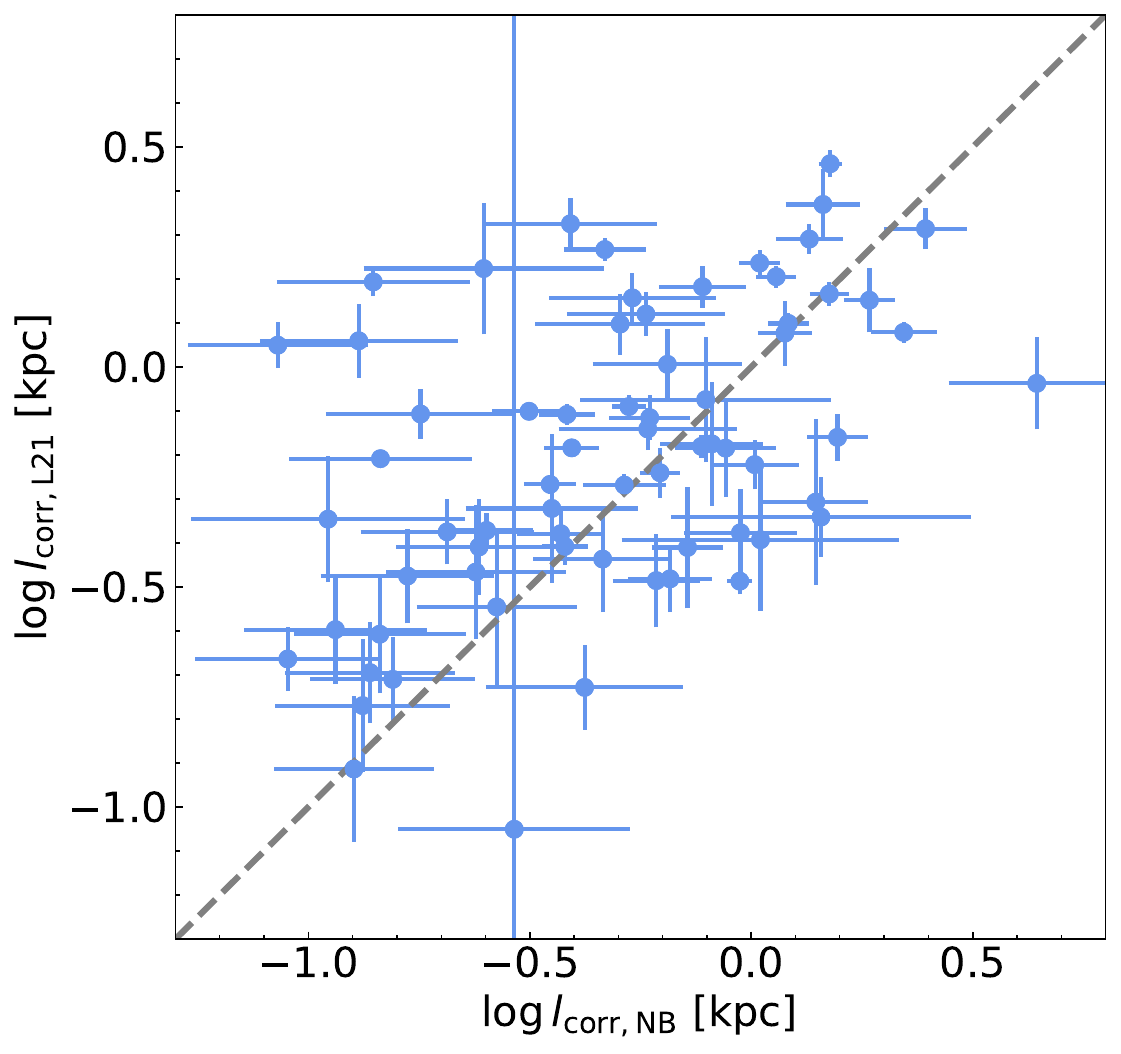}}
    \caption{Comparison between the correlation lengths calculated by \citetalias{li2021} from metallicity maps produced using the [N~\textsc{ii}]/[O~\textsc{ii}] metallicity diagnostic from \citet{kewley2019a} (vertical axis) and the correlation lengths estimated in this work using \textsc{NebulaBayes} (horizontal axis). Each point corresponds to one of the 68 galaxies in common to both samples; points show median values, while error bars show the 16th to 84th percentile range. The dashed grey line indicates the 1 to 1 relation.}
    \label{fig:l_corr_campare}
\end{figure}

\begin{figure*}
    \resizebox{17cm}{!}{\includegraphics{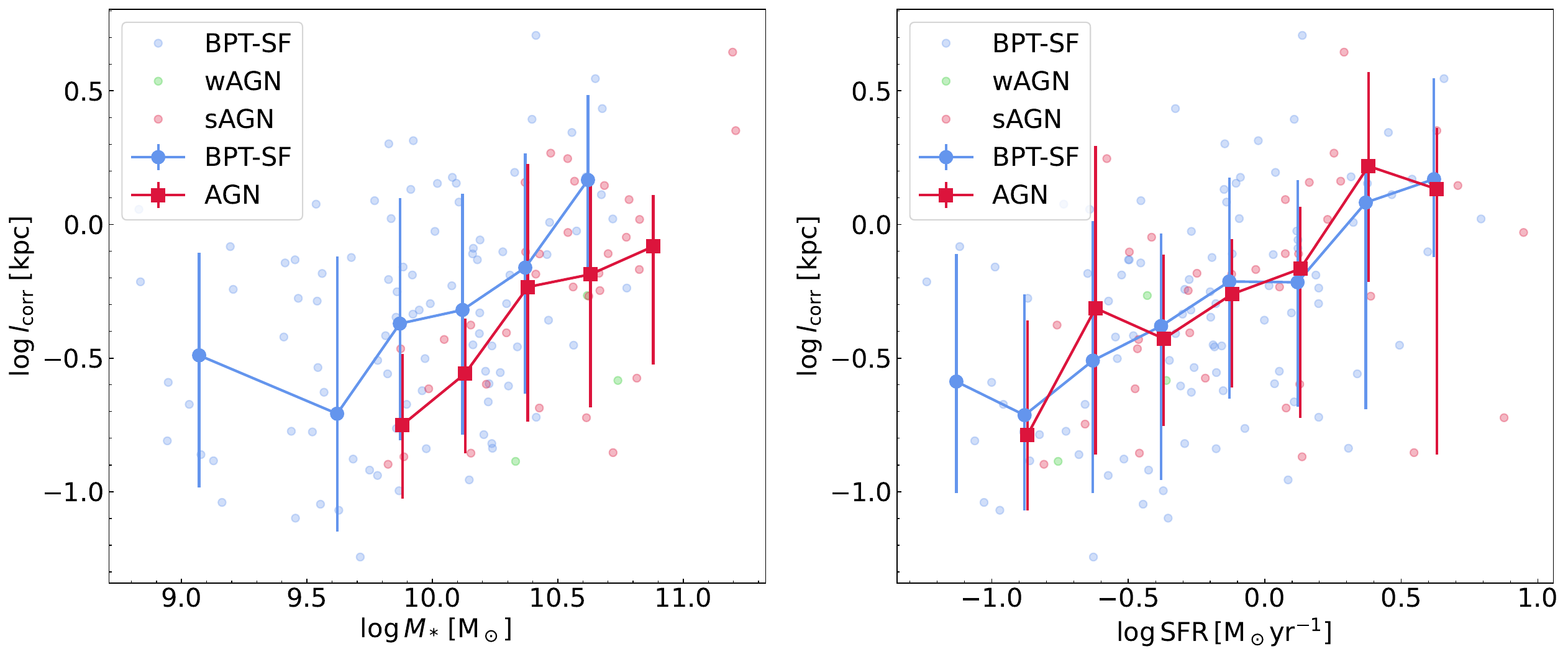}}
    \caption{Left: $\log \lcor$ versus  $\log M_*$ for BPT-SF (blue), wAGN (green), and sAGN (red) galaxies. The blue circles with error bars represent the median and 16th to 84th percentile range of $\lcor$ for BPT-SF galaxies in each stellar mass bin, computed via the procedure described in \autoref{sec:results}; red squares with error bars show the same quantities for AGN-host galaxies, combining the wAGN and sAGN groups. The stellar mass bins are the same for both BPT-SF and AGN-host galaxies, but the points have been offset slightly for readability. Right: same as the left panel, but showing $\log\lcor$ versus $\log \mathrm{SFR}$ rather than versus $\log M_*$.}
    \label{fig:l_corr_relation}
\end{figure*}

Before we examine how galaxy correlation lengths differ between AGN-host and non-AGN galaxies, and depend on other galactic properties in both populations, we firstly verify that our correlation lengths measurements are reliable and not subject to the influences of the spatial resolution of CALIFA. Following \citetalias{li2021}, \autoref{fig:beam} shows the estimated correlation lengths, $\lcor$, as a function of the beam size $\sigma_\mathrm{PSF} (1+\cos i)/(2\cos i)$ in physical scales (kpc) for the 96 galaxies with PSF measurements in CALIFA. The Pearson correlation coefficient for the BPT-SF galaxies is 0.27, similar to what found in \citetalias{li2021} and indicating that our estimates of $\lcor$ are at most weakly dependent on spatial resolution. One might be concerned that the measured $\lcor$ are comparable to the beam size, but note that the two-point correlation function contains significant structure even at sizes of several times $l_\mathrm{corr}$ (c.f.~Figure 1 of \citealt{krumholz2018}), and thus that observations can constrain $l_\mathrm{corr}$ even if the inner structure of the two-point correlation function on scales $< l_\mathrm{corr}$ is masked by beam smearing.

While the BPT-SF sample shows only a weak correlation between beam size and $\lcor$, the Pearson correlation coefficient of AGN-host galaxies (sAGN and wAGN) is 0.55, showing a stronger correlation that is potentially of concern. Before concluding that this is a beam-smearing effect, however, we must consider an alternative explanation: due to CALIFA's fixed bundle size, the survey selects galaxies of roughly fixed angular size, and as a result the galaxies it includes at larger distances tend to have larger physical radii and higher masses. Since \citetalias{li2021} show that more massive galaxies in turn have larger $\lcor$, we would therefore expect galaxies with larger PSFs -- which are on average further away and more massive -- to have larger $\lcor$ even in the absence of beam effects.

To understand whether the correlation we find in \autoref{fig:beam} between beam size and $\lcor$ is due to this innocuous effect or to beam smearing, we carry out a partial-correlation test between stellar mass, beam size, and $\lcor$ for AGN-host galaxies. When controlling for beam size,  we find that the partial correlation coefficient between stellar mass and $\lcor$ is 0.46 with a $p$-value of 0.02; by contrast, when controlling for stellar mass the partial correlation coefficient between beam size and $\lcor$ is 0.22 with a $p$-value of 0.26, strongly suggesting that the correlation between $l_\mathrm{corr}$ and beam size in the AGN sample is a byproduct of CALIFA's sample selection, not an indication of beam-smearing effects.

For completeness we also repeat the partial-correlation test for BPT-SF galaxies. When controlling for beam size in this sample the partial correlation coefficient between stellar mass and $\lcor$ is 0.26 with a $p$-value of 0.04, while the partial correlation coefficient between beam size and $\lcor$ controlling for stellar mass is 0.12 with a $p$-value of 0.33. This suggests that the minor correlation between $\lcor$ and beam size that we find in the BPT-SF sample is also primarily due to the correlation between $\lcor$ and stellar mass interacting with CALIFA's selection function, and not to beam-smearing.

Having shown that our results of $\lcor$ are not strongly influenced by CALIFA's resolution, we next verify that our correlation lengths are in reasonable agreement with those derived by earlier authors. \autoref{fig:l_corr_campare} compares the $\lcor$ values we derive here with those obtained by \citetalias{li2021} for the 68 galaxies common to both samples (which in the case of \citetalias{li2021} includes only BPT-SF galaxies, since they were required to exclude AGN). \citetalias{li2021} obtained their metallicities using the [N~\textsc{ii}]/[O~\textsc{ii}] metallicity diagnostic from \citet{kewley2019a}. As the figure shows, the results from the two studies generally follow the 1 to 1 relation with a 1$\sigma$ scatter of $\approx 0.34$ dex; the Pearson correlation coefficient between the two data sets is 0.49. \citetalias{li2021} compared the correlation lengths of maps derived using three different line-ratio metallicity diagnostics and found similar scatter and Pearson correlations when comparing them. We thus confirm that our $\lcor$ estimates are consistent with those obtained by \citetalias{li2021} to the extent that we expect given that we are using different metallicity diagnostics.

\subsection{Correlation lengths in AGN-host versus BPT-SF galaxies}

\autoref{fig:l_corr_relation} shows $\lcor$ versus stellar mass (left panel) and SFR (right panel) for both BPT-SF (non-AGN, blue) and AGN-host (combining the wAGN and sAGN groups, red) galaxies. Points show the median over all MCMC samples for $\lcor$ (excluding the burn-in period) for all galaxies falling into a given bin of stellar mass or SFR, while error bars indicate the 16th to 84th percentile range over these samples; for both stellar mass and SFR, bins are 0.25 dex wide. For both BPT-SF and AGN-host galaxies, we find $\lcor$ increases with both stellar mass and SFR. These positive correlations are similar to those found in \citetalias{li2021} and \citetalias{li2023}.

However, while BPT-SF and AGN-host galaxies show trends in the same direction, we also find that there is a systematic offset between the two galaxy groups at fixed stellar mass -- AGN-host galaxies have systematically lower $\lcor$ than BPT-SF galaxies of similar stellar mass. There is no corresponding systematic difference at fixed SFR. To verify that this visual impression is statistically robust, we carry out $t$-tests comparing the distributions of $\lcor$ values for BPT-SF and AGN-host galaxies over the four bins of stellar mass where the two samples overlap (the left panel of \autoref{fig:l_corr_relation}). The null hypothesis of this test is that the mean $\lcor$ for AGN-host galaxies is greater than or equal to that of the BPT-SF galaxies in the same mass bin. The $p$-values returned by this test are, from lowest to highest stellar mass bin, 0.020, 0.111, 0.327, and 0.007. If we take $p=0.05$ to mark significance then the results rule out the null hypothesis in two of the four bins. \footnote{Note that the apparent difference between the BPT-SF and AGN-host samples in the bin $\log (M_*/\Msun) \in [10.25, 10.5)$ does not reach statistical significance because there are only four AGN galaxies in this stellar mass bin.} Repeating this exercise for the two galaxy classes binned by SFR with the null hypothesis that the mean $\lcor$ for AGN-host galaxies is equal to that of BPT-SF galaxies in the same SFR bin (right panel of \autoref{fig:l_corr_relation}), on the other hand, yields $p$-values of 0.80, 0.43, 0.85, 0.76, 0.82, 0.22, and 0.48 from the lowest to the highest $\log \mathrm{SFR}$ bin, consistent with there being no statistically-significant difference in the distributions.

Comparing \autoref{fig:l_corr_relation} to \autoref{fig:sfms} yields the interesting conclusion that the offset between BPT-SF and AGN-host galaxies in the $\lcor$ versus $M_*$ relation is qualitatively very similar to the offset between the two samples in SFR versus $M_*$. That is, at fixed stellar mass, galaxies that host AGNs have both systematically smaller SFRs and metallicity $\lcor$, and the direction and size of the offset are similar in both diagrams. Consequently, there is no offset between the two galaxy classes at fixed SFR, and thus we can attribute the full difference in $\lcor$ values between BPT-SF and AGN-host galaxies at fixed stellar mass to differences in SFR.

\section{Discussion}
\label{sec:discussion}

In this section, we explore the implications of our results. We firstly discuss the implications of our findings for how AGN affect galaxy properties in \autoref{sec:agn_influence}, and then discuss the implications for what processes regulate galactic metallicity distributions in \autoref{sec:fund_para}.

\subsection{Implications for how AGN influence galaxy properties}
\label{sec:agn_influence}

We have found that AGN-host galaxies have smaller $\lcor$ than BPT-SF galaxies (non-AGN galaxies) at fixed stellar mass (the left panel of \autoref{fig:l_corr_relation}), but that this result is fully accounted for by the smaller SFR of AGN hosts at fixed stellar mass (\autoref{fig:sfms}), such that there is no difference in $\lcor$ between BPT-SF and AGN-host galaxies of similar SFR (right panel of \autoref{fig:l_corr_relation}). This suggests that AGN activity does not directly influence galactic metallicity distributions, but instead only affects them indirectly by suppressing star formation or at least by correlating with low SFR.

At first this result is surprising. In the context of the \citet{krumholz2018} model, the correlation length is expected to be $\lcor = \sqrt{\kappa t_*}$, where $\kappa$ is a diffusion coefficient describing diffusion of metals in the ISM and $t_*$ is a characteristic timescale over which star formation has taken place. The diffusion coefficient in turn should be related to the velocity dispersion $\sigma_g$ and scale height $h$ of the ISM as $\kappa \approx h\sigma_g /3$ (\citealt{karlsson2013}; \citetalias{li2023}). Any physical processes that enhances the strength of the turbulence should produce an increase in $\sigma_g$, and thus an increase in the $\lcor$. Powerful AGN like quasars are able to trigger strong outflows and shocks and produce large velocity dispersions \citep[e.g.,][]{veilleux2005,rich2015,davies2017,oh2022}, and thus we might expect AGN activity to increase $\lcor$. Moreover, AGN activity has long been thought to be correlated with galactic mergers \citep[e.g.,][]{hopkins2008,koss2018,comerford2024}, and merging systems are also observed to have large $\lcor$ \citepalias{li2021, li2023}. Considering these effects, one might expect that AGN-host galaxies possess larger $\lcor$ than non-AGN galaxies. However, as discussed above, we find the opposite.

We suspect this is because in local universe, most AGN are not powerful enough to trigger strong galactic-scale winds or shocks. As the example sAGN shown in \autoref{fig:bpt_example} demonstrates, even sAGN in our sample have AGN excitation restricted to the central regions of the galaxy, while the bulk of the disc is still ionised by massive stars. Quantitatively, in \aref{app:sfr_corr} we show that even in AGN-host galaxies the AGN typically provides only $\sim 1\%$ of the total ionisation. \citet{lopez2019,lopez2020} find that in both the CALIFA and AMUSING++ samples only a small fraction of AGN hosts harbour detectable outflows. Together all this evidence suggests that most AGN in the local Universe are not capable of substantially altering the bulk properties of the ISM. Similarly, while mergers may be important triggers of quasars and similar very bright AGN activity, external perturbations may contribute only marginally to the sorts of low-luminosity AGN present in our sample; these may instead be triggered predominantly by local secular processes occurring in the galactic nucleus, with no substantial contribution from the types of mergers and interactions that can raise $\lcor$ \citep{orban2011,kocevski2012}.

To the extent that this hypothesis is correct and our AGN are too dim to affect ISM properties directly via shocks or to be correlated with mergers, an alternative effect arising from AGN can be dominant. This is that optically-selected AGN have lower SFR than galaxies on the star-forming main sequence \citep{schawinski2014,ellison2016,lacerda2020}. This correlation in turn can arise through two possible channels. One is negative AGN feedback that takes the form not of violent ejection of mass from the ISM, which presumably would be sufficient to affect metallicity distributions directly \citep[e.g.,][]{hopkins2008,hopkins2010,bing2019}, but via a more gentle path of preventing ongoing gas accretion, namely ``starvation'' \citep{kumari2021}. The other is that AGN activity is associated with bulge growth \citep{heckman2004,kormendy2013,sanchez2018}, which inhibits star formation (so-called ``morphological'' quenching) due to increased shear stabilising the gaseous components of galactic discs \citep{martig2009,fang2013}. In either of these scenarios, AGN would affect galaxy metallicity distributions only indirectly, via the resulting reduction in the SFR, consistent with what we see in the observations. 

Our finding that the offset between AGN-hosts and SF galaxies at fixed stellar mass is due to differences in SFR is also consistent with the results of \citetalias{li2024}, who find that AGN hosts generally have higher metallicty than BPT-SF galaxies at fixed stellar mass, but that this difference is a natural result of the existence of the FZR, whereby the SFR anti-correlates with the metallicity at fixed stellar mass. Since AGN hosts have smaller SFRs at fixed stellar mass, the FZR implies that they will have larger metallicities, an indirect effect directly analogous to the one we have found here. Taken together, the two results suggest that, at least for low-luminosity AGN, the primary way that AGN affect the gas-phase metallicities of galaxies -- both the total metallicity and the metallicity distribution -- is that hosting an AGN is associated with galaxies having lower SFRs, rather than via any direct effects of the AGN itself.

\subsection{Which galactic parameter is most important to regulating metallicity distributions?}
\label{sec:fund_para}

Both \citetalias{li2021} and \citetalias{li2023} find that galaxies' $\lcor$ is positively correlated with their stellar mass, SFR, and size. However, because each of these quantities correlates with the other two, they are not able to determine which of these three parameters is the fundamental one when it comes to setting $\lcor$, and which are secondary correlations. Our current sample offers the opportunity to break this degeneracy: the lower $\lcor$ of AGN hosts compared to non-AGN at fixed stellar mass but similar $\lcor$ at fixed SFR strongly suggests that SFR is more fundamental than stellar mass when it comes to regulating $\lcor$, and thus galaxies' internal metallicity distributions more broadly. To quantify this intuition, we carry out a partial-correlation test among $\lcor$, stellar mass, and SFR for BPT-SF galaxies with the {\tt pingouin} python package. When controlling the stellar mass, the partial correlation coefficient between $\lcor$ and SFR is 0.21 with a $p$-value of 0.04; by contrast, controlling for SFR the partial correlation coefficient between $\lcor$ and stellar mass is 0.08 with the $p$-value of 0.42. This test is consistent with the result suggested by the comparison of the BPT-SF and AGN-host samples, which is that SFR is the more important parameter. This is also consistent with theoretical expectation: there is a well-known and theoretically-expected correlation between galaxies' SFR and velocity dispersions \citet[][and references therein]{krumholz2018b}, and to the extent that larger velocity dispersion yields larger diffusion coefficients, this provides a natural pathway by which an increase in SFR could increase $\lcor$.

As in \autoref{sec:agn_influence}, we can make a useful connection to the results of \citetalias{li2024} for total metallicity. The gas-phase metallicity primarily records the cumulative effects of metal enrichment via stellar nucleosynthesis in a galaxy through cosmic time \citep{asari2009}. If we take stellar mass as representing an evolutionary sequence, the mass-metallicity relation then reflects the cumulative buildup of metals over cosmological time scales \citep[e.g.,][]{maiolino2008,zahid2011}. However, on top of this secular trend, the metallicity also responds to processes like star formation that can fluctuate on much shorter timescales, and the anti-correlations between SFR (or $\Sigma_\mathrm{SFR}$) and metallicity at fixed stellar mass (or $\Sigma_*$) are the signature of this response to short timescale phenomena \citep[e.g.,][]{lara2010,mannucci2010,baker2023}. In this context, our findings here present another example of this short timescale response, this time in the spatial structure of fluctuations in metallicity fields rather than in the mean metallicity.

\section{Conclusion}
\label{sec:concluciton}

In this work we measure the two-point correlation functions of the metallicity fields of 95 galaxies that do not contain optical AGN and 37 AGN-host galaxies drawn from CALIFA IFS survey. The main methodological innovation in our work is that we measure metallacities using a Bayesian method (\textsc{NebulaBayes}) that self-consistently accounts for the AGN contribution to emission-line fluxes spaxel-by-spaxel, which allows us to derive metallicities for the AGN-host and non-AGN galaxies using the same method and without incurring the large systematic uncertainties that are inevitable when comparing metallicities derived using different diagnostics. This in turn makes a direct comparison between AGN-host and non-host galaxies possible, allowing us to achieve our primary science aim: understanding how AGN influence galactic metallicity distributions.

We find that the two-point correlation functions of both AGN host and non-host galaxies are well fit by the simple metal injection and diffusion model proposed by \citet{krumholz2018}. The key fitting parameter we can derive from this model is correlation length $\lcor$, which describes the characteristic length scale over which metallicity fluctuations are correlated. Our results for non-AGN host galaxies show reasonable consistency with earlier results from \citet{li2021} derived from metallicity maps obtained from traditional emission-line flux ratio diagnostics, validating our new Bayesian methods.

Comparing $\lcor$ between non-AGN and AGN-host galaxies, we find that AGN hosts generally have smaller $\lcor$ than non-host galaxies at fixed stellar mass (the left panel of \autoref{fig:l_corr_relation}), but similar $\lcor$ at fixed SFR (the right panel of \autoref{fig:l_corr_relation}). We show that the difference in $\lcor$ at fixed stellar mass is fully explained by the fact that AGN-host galaxies have reduced SFR at fixed stellar mass compared to non-hosts (\autoref{fig:sfms}). This finding suggests that AGN, at least over the luminosity range we are able to probe, influence the metallicity distributions within galaxies only indirectly, either by suppressing star formation through starvation or because AGN grown is correlated with bulge growth, which suppresses star formation through morphological quenching. We find no evidence for other, more direct effects. The fact that the two galaxy-subgroups have similar $\lcor$ at fixed SFR, but not at fixed stellar mass, further suggests that SFR is more fundamental than stellar mass when it comes to regulating galaxy metallicity distributions. Partial-correlation tests among the $\lcor$, stellar mass, and SFR confirm this impression. This suggests that, while galaxy mean metalicities are the result of accumulation over the full star formation history of galaxies, fluctuations in the metallicity distribution within galaxies are more driven by short-term responses to physical processes such as star formation that can change much faster than a Hubble time.

\section*{Acknowledgements}

We thank the referee Prof. Roberto Maiolino for a detailed report that helped significantly in improving the presentation of our work. This study uses data provided by the Calar Alto Legacy Integral Field Area (CALIFA) survey (http://califa.caha.es/). This study is based on observations collected at the Centro Astron\'omico Hispano Alem\'an (CAHA) at Calar Alto, operated jointly by the Max-Planck-Institut f\"ur Astronomie and the Instituto de Astrof\'isica de Andaluc\'ia (CSIC). This study also used computational resources provided by the National Computational Infrastructure, which is supported by the Australian Government, through award jh2. MRK acknowledges support from the Australian Research Council through its Laureate Fellowship funding scheme, award FL220100020. SFS thanks the PAPIIT-DGAPA AG100622 project and CONACYT grant CF19-39578. We thank Trevor Mendel for comments and ideas for this work.

\section*{Data Availability}

The data underlying this article will be shared on reasonable request with the corresponding author.



\bibliographystyle{mnras}
\bibliography{Li24}



\appendix

\section{Potential biases due to AGN-host identification method}
\label{app:heii}

\begin{figure}
{\includegraphics[width=\columnwidth]{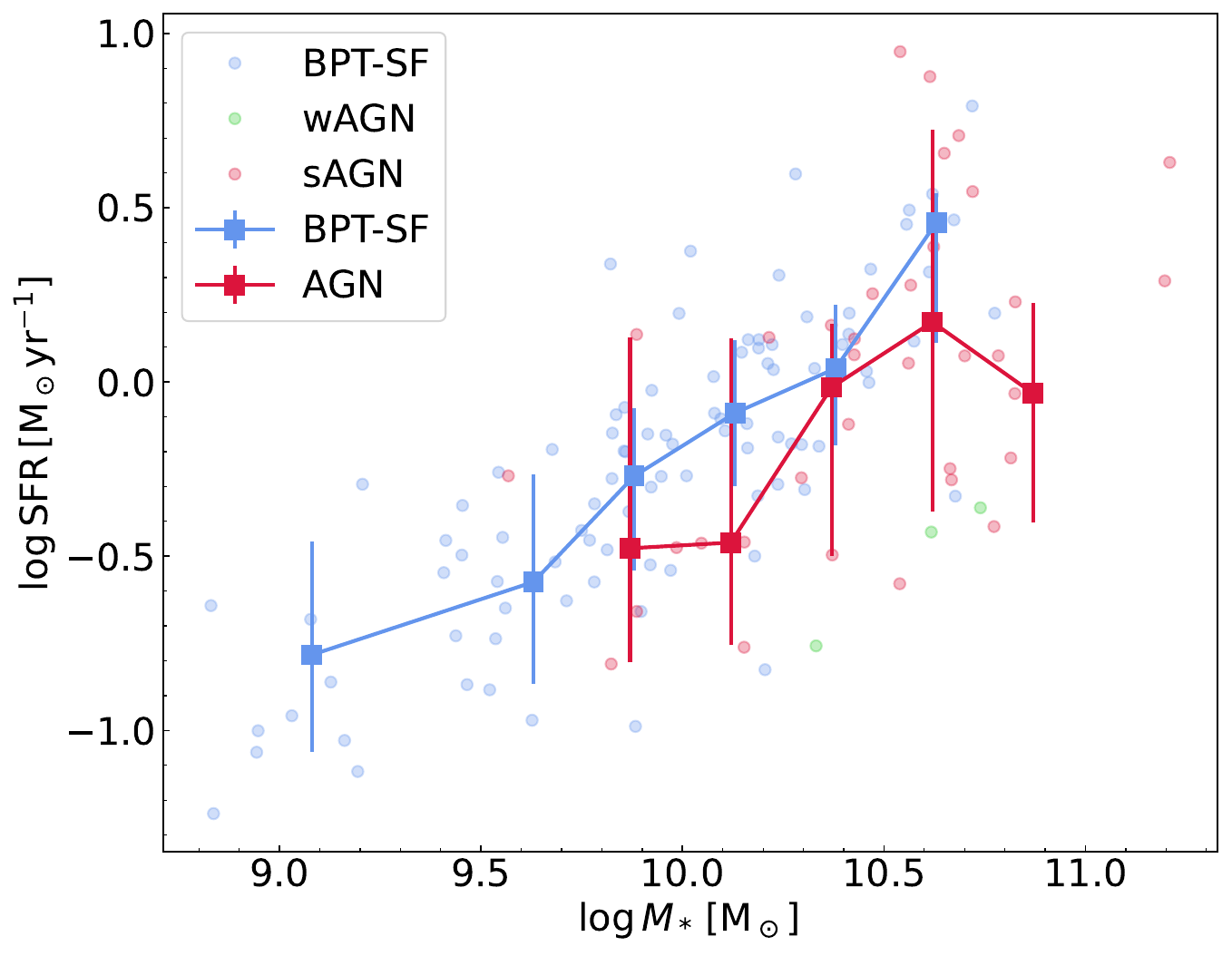}}
    \caption{Same as \autoref{fig:sfms}, but with two galaxies re-classified from BPT-SF to sAGN based on the He~\textsc{ii} diagram as discussed in \aref{app:heii}.}
    \label{fig:sfms_heii}
\end{figure}

\begin{figure*}
    \resizebox{17cm}{!}{\includegraphics{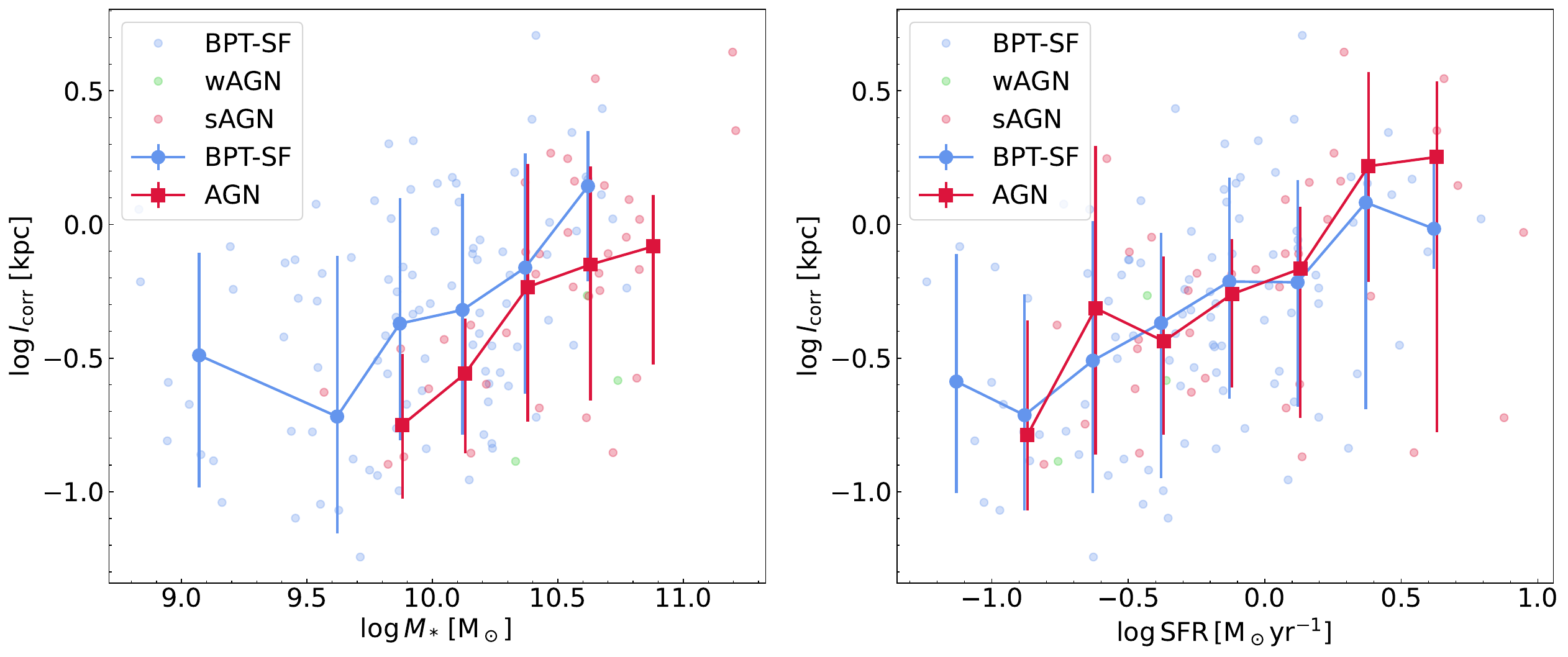}}
    \caption{Same as \autoref{fig:l_corr_relation}, but with two galaxies re-classified from BPT-SF to sAGN based on the He~\textsc{ii} diagram as discussed in \aref{app:heii}.}
    \label{fig:l_corr_relation_heii}
\end{figure*}

As discussed in \autoref{sec:sample}, we introduce a potential bias by selecting our AGN-host sample based on the BPT diagram, since this may selectively miss AGN-hosts that have higher SFRs and thus stellar-driven line emission bright enough to mask weak AGN signatures. \citet{tozzi2023} explore this possibility using the MaNGA sample; they compare AGN identified using the traditional BPT diagram to those identified based on a modified version of the BPT diagram with [O~\textsc{iii}]/H$\beta$ replaced by He~\textsc{ii}$~\lambda 4686$/H$\beta$. The He~\textsc{ii} line is more sensitive to AGN and much less likely to be masked by bright stellar-driven nebular emission due to its higher excitation energy (54.4 eV versus 35.2 eV for [O~\textsc{iii}]). \citeauthor{tozzi2023} find that 6\% of AGN-host galaxies can be identified only using the He~\textsc{ii} line, and that these additional AGN-host galaxies all lie on the star-forming main sequence and have SFRs similar to those of BPT-SF galaxies.

To test for this effect in our sample, we stack the flux of He~\textsc{ii}, H$\beta$, H$\alpha$, and [N~\textsc{ii}] within central $2\farcs50$ of each galaxy, which is the beam size of CALIFA, and adopt the same SNR cut and the same demarcation line to classify AGN as in \citet{tozzi2023}. We find that seven galaxies out of the 132 in our sample are identified as AGN hosts by this method, but that only two of these were classified as BPT-SF based on the traditional BPT diagram. If we re-classify these extra two galaxies as AGN, our AGN sample grows from 37 to 39, an increase of 6\%, completely consistent with \citeauthor{tozzi2023}'s result that 6\% of AGN can be found using the He~\textsc{ii} line only.

We now repeat the analysis in the paper for a modified sample in which we re-classify these two galaxies from BPT-SF to sAGN in order to determine how our results change. \autoref{fig:sfms_heii} shows the SFR versus stellar mass relation of AGN hosts and BPT-SF galaxies after this re-classification. Comparing with \autoref{fig:sfms}, we see that our qualitative result that AGN hosts have generally smaller SFRs than BPT-SF galaxies at fixed stellar mass remains unchanged. Similarly, \autoref{fig:l_corr_relation_heii} shows the relationship between $\lcor$, stellar mass, and SFR we obtain after this re-classification. Comparing this diagram with \autoref{fig:l_corr_relation}, the most obvious difference is in the bin $\log \mathrm{SFR} \in [0.5, 0.75]$, where now the median $\lcor$ of AGN hosts is apparently greater than that of BPT-SF galaxies. However, this visual shift is mostly driven by the very small number of AGN in this bin, and is not statistically significant: a $t$-test with the null hypothesis that the mean $\lcor$ of BPT-SF is the same as that of AGN hosts returns a $p$-value of 0.98. Our main conclusions therefore again remain unchanged.

\section{Influences of the priors of the beam size}
\label{app:sigma_beam}

\begin{figure*}
    \resizebox{17cm}{!}{\includegraphics{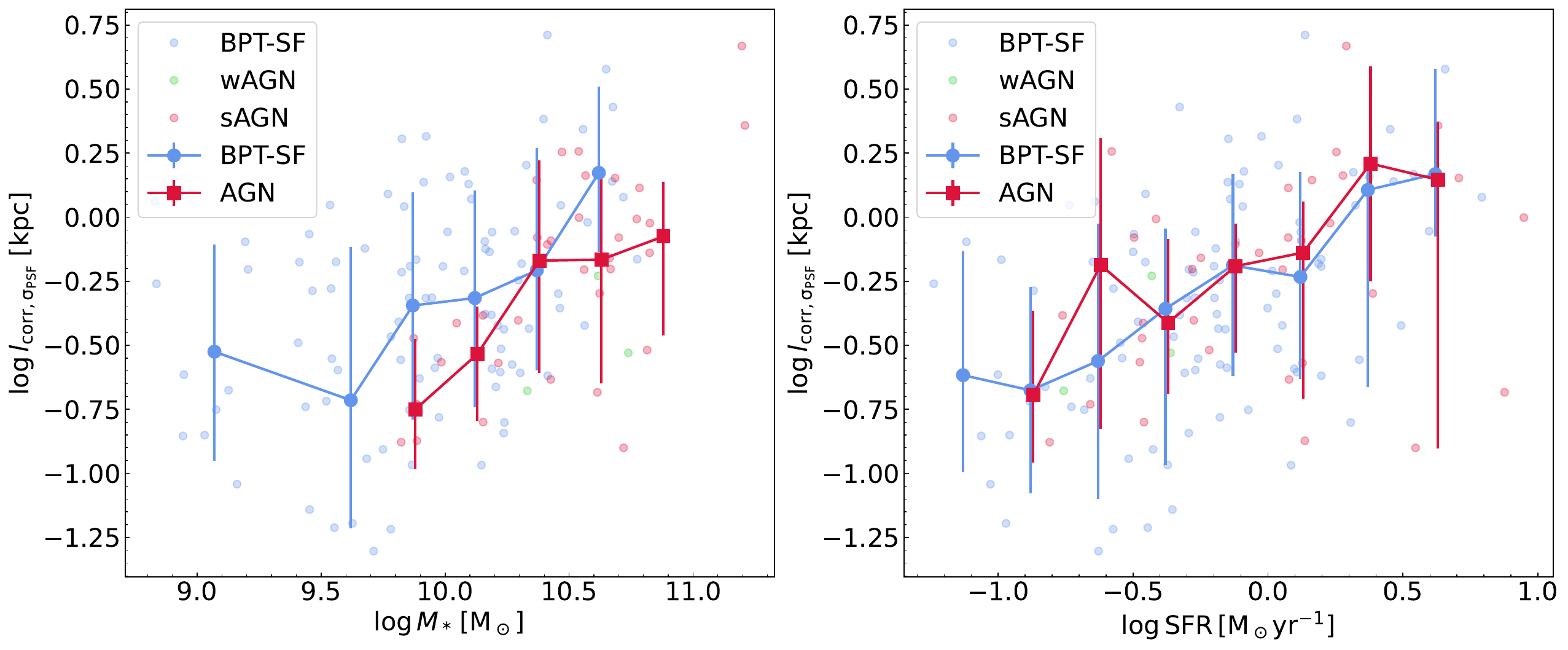}}
    \caption{Same as \autoref{fig:l_corr_relation}, but for $\lcor$ derived from a fit where we set the prior on $\mathrm{\sigma_{beam}}$ to a very narrow Gaussian centred on $\mathrm{\sigma_{PSF}}$, the measured PSF size from CALIFA; see text for details.}
    \label{fig:l_corr_relation_beam}
\end{figure*}

\begin{figure*}
    \resizebox{17cm}{!}{\includegraphics{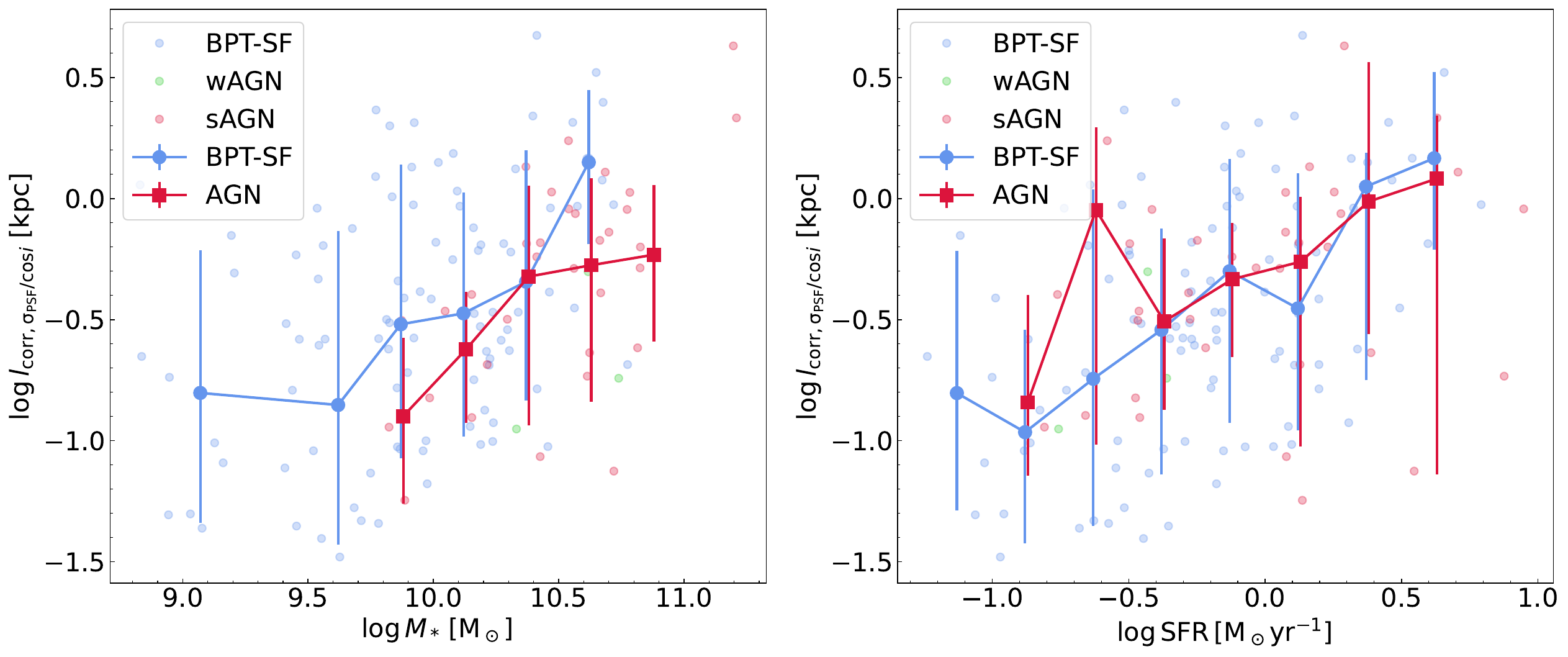}}
    \caption{Same as \autoref{fig:l_corr_relation_beam}, but now setting the central value of the prior on beam size to $\sigma_\mathrm{PSF}/\cos i$, where the $\mathrm{\sigma_{PSF}}$ is the measured PSF size from CALIFA and $i$ is the inclination angle calculated from \autoref{equ:cosi}.}
    \label{fig:l_corr_relation_deproject}
\end{figure*}

As discussed in \autoref{sec:lcorr}, due to deprojection of the metallicity maps the observational beam along the minor axis of the galaxies is stretched by a factor of $1/\cos i$, where $i$ is the inclination angle calculated from \autoref{equ:cosi}. In the main text we handle this issue by setting our prior on the beam size, $\sigma_\mathrm{beam}$, to a Gaussian whose centre is the arithmetic mean of $\sigma_\mathrm{PSF}$ and $\sigma_\mathrm{PSF}/\cos i$, the beam sizes along the major and minor axis, and with a width equal to half the difference between the major and minor axis beam sizes. In this appendix we test the sensitivity of our results to this choice by repeating the full analysis pipeline in the main paper using two extreme alternatives for the prior.

\autoref{fig:l_corr_relation_beam} and \autoref{fig:l_corr_relation_deproject} show the relation between $\lcor$ and stellar mass as well as the relation between $\lcor$ and SFR for both AGN-host galaxies and BPT-SF galaxies that we derive by setting the mean of the Gaussian prior on the $\sigma_\mathrm{beam}$ to $\sigma_\mathrm{PSF}$ and $\sigma_\mathrm{PSF}/\cos i$, respectively; in both cases we set the standard deviation on the prior to of $0\farcs 01$, a very small value that essentially fixes $\sigma_\mathrm{beam}$ to the value we assign for the mean. Again, for those without a PSF measurement, we assign the mean of the Gaussian prior on the $\sigma_\mathrm{beam}$ to $1\farcs06$ and $1\farcs06/\cos i$, and corresponding standard deviation to $0\farcs14$ and $0\farcs14/\cos i$, respectively, where $1\farcs06$ and $0\farcs14$ are the mean and the $1\sigma$ level of scatter of the PSF distribution in CALIFA, respectively. We are therefore testing how our results change if we force the beam sizes to the largest and smallest plausible values. Comparing these two diagrams with \autoref{fig:l_corr_relation}, we find a qualitatively similar result that AGN hosts have smaller $\lcor$ than BPT-SF galaxies at fixed stellar mass, but show no apparent difference at fixed SFR. We thus confirm that our main conclusion remain unchanged even under extreme assumptions about beam sizes.

\section{Failed fits to the two-point correlation function}
\label{app:fail}

\begin{figure*}
    \resizebox{17cm}{!}{\includegraphics{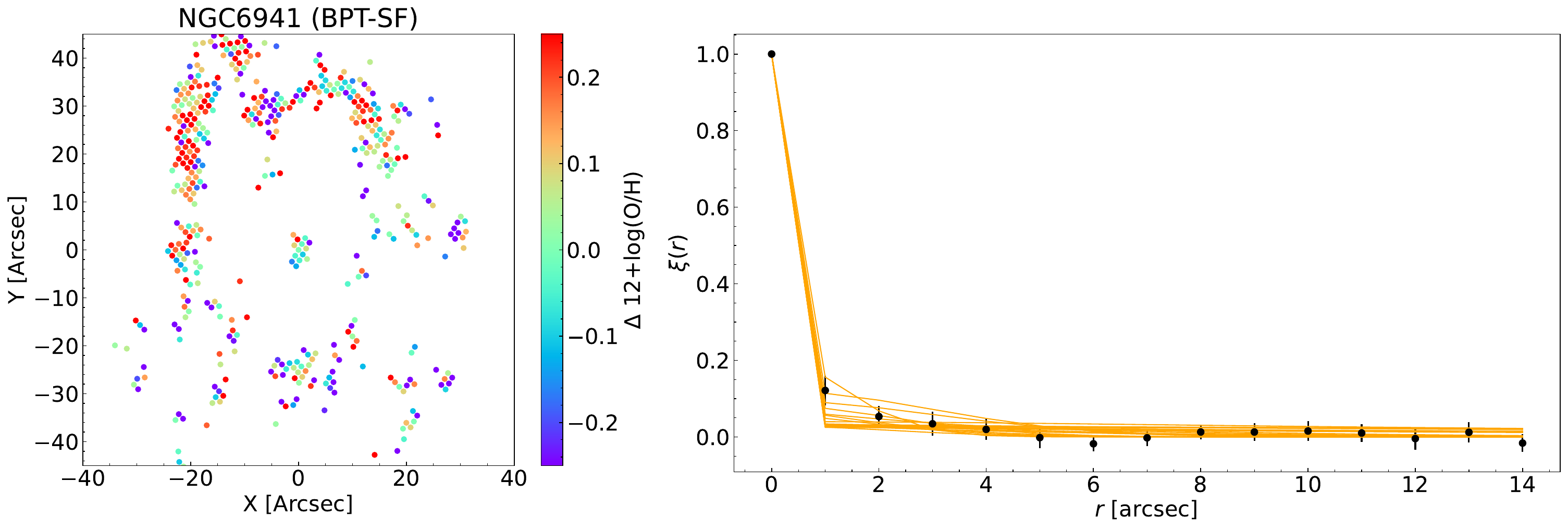}}
    \caption{The same as \autoref{fig:corr_example}. However, we show an example of which we fail to fit the two-point correlation function with the procedures elaborated in \autoref{sec:lcorr}.}
    \label{fig:fail_example}
\end{figure*}

\begin{figure}
{\includegraphics[width=\columnwidth]{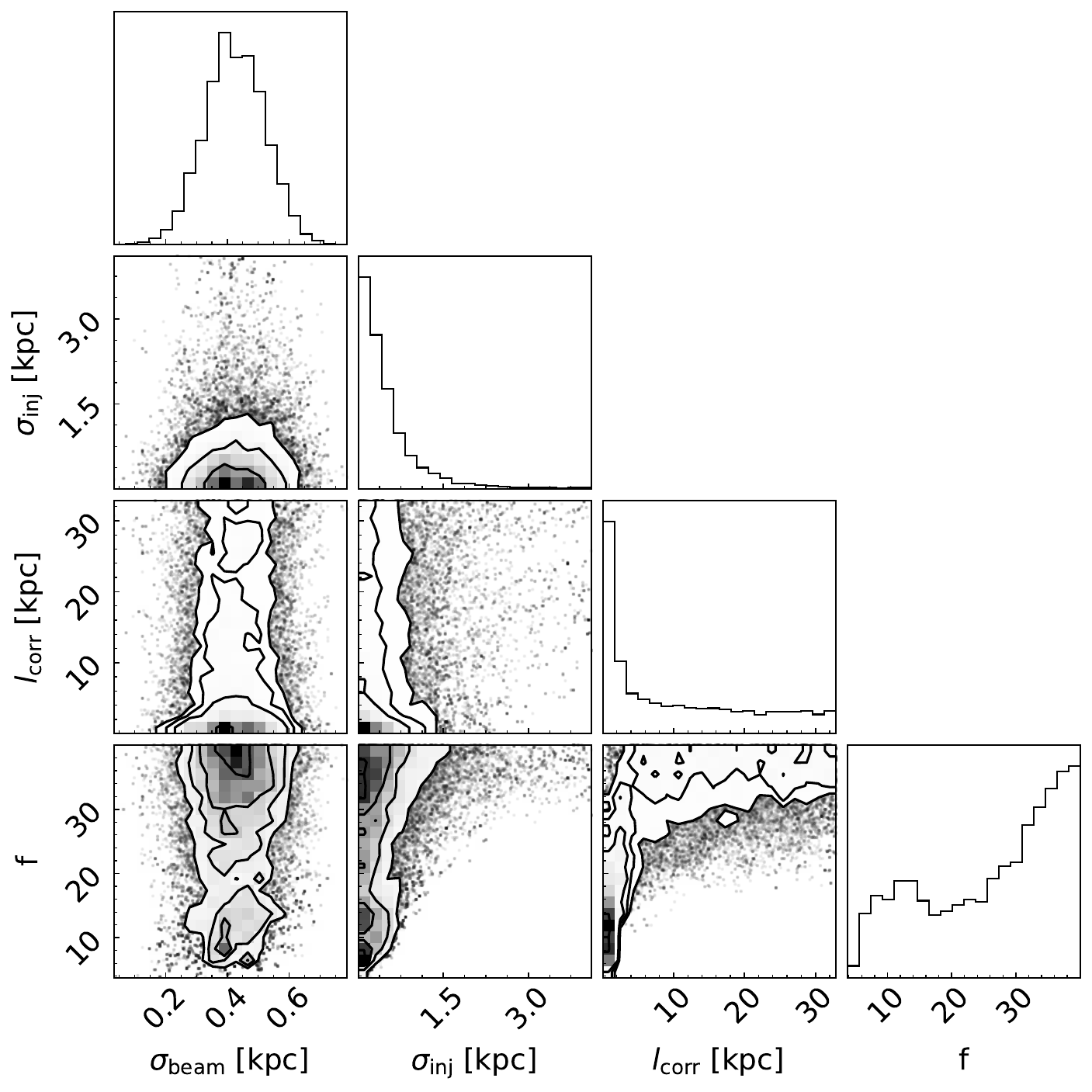}}
    \caption{The same as \autoref{fig:tri_example}. But we show the posterior distribution of four parameters for NGC6941, a galaxy for which our MCMC fitting does not produce a good measurement of $\lcor$.}
    \label{fig:triangle}
\end{figure}

By inspecting the posterior PDF of the MCMC sampling, we find that we fail to obtain a reasonable $\lcor$ for three galaxies: NGC4470, NGC1070, and NGC6941. For each of these galaxies the posterior PDF of $\lcor$ extends up to the maximum allowed by our prior, and is in general extremely broad, suggesting that our MCMC is not finding an acceptable model. To explore the origin of this failure, in \autoref{fig:fail_example} and \autoref{fig:triangle} we show the metallicity fluctuation map, two-point correlation function, and posterior PDF of the parameters describing the two-point correlation function for one of these failed galaxies NGC6941; the other two are qualitatively similar, and the explanation we present in this Appendix applies equally well to them.

We first see that for this galaxy the metallicity fluctuation map is sparsely sampled due to the SNR cut described in \autoref{sec:sample}. To understand if this is the primary reason for the failed fit, we calculate the area filling factors within the convex hull following the procedure described in \citetalias{li2023}. We find filling factors of 0.17 for NGC6941, 0.23 for NGC1070, and 0.35 for NGC4470, all of which are larger than the threshold of 0.04 obtained empirically by \citetalias{li2023} for their sample. Moreover, while these filling factors are somewhat low, there are a large number of galaxies in our sample that we do successfully fit despite their similar or even smaller filling factors.

Instead, we can understand the origin of the failure by noting that the joint posterior PDF of $f$ and $\lcor$ in \autoref{fig:triangle} shows two ``wings'', one with $\lcor$ smaller than the beam size and smaller $f$ and another with large $f$ and essentially a uniform distribution of $\lcor$. We can understand the origin of these two branches by examining the two-point correlation function to which we are fitting (right panel of \autoref{fig:fail_example}). The essential point to notice is that this galaxy shows essentially no correlations in its metallicity field -- excluding the point at zero lag which is always unity, the most correlated bin only has a central value $\xi\approx 0.1$, and given the size of the error bar this point is consistent with $\xi = 0$ at the $2\sigma$ level. In terms of the model to which we are fitting, there are two ways to produce a two-point correlation function that looks like the right panel of \autoref{fig:fail_example}. One is if the correlation length is significantly smaller than the size of the beam, so that no correlations exist on resolved scales, and this possibility corresponds to the small $\lcor$ wing of the posterior PDF. The other is if the observational error in the metallicity, parameterised by $f$, is so large that any correlations present in the data are buried below the noise; this corresponds to the large $f$ wing of the posterior PDF, for which the marginal posterior PDF of $\lcor$ is simply the prior, because the data are too noisy to permit the correlation to be measured.

Given this understanding of the fit failures, we conclude that we cannot meaningfully constrain $\lcor$ for these three galaxies, and we therefore exclude them from our analysis.

\section{On AGN corrections to the SFR}
\label{app:sfr_corr}

\begin{figure}
{\includegraphics[width=\columnwidth]{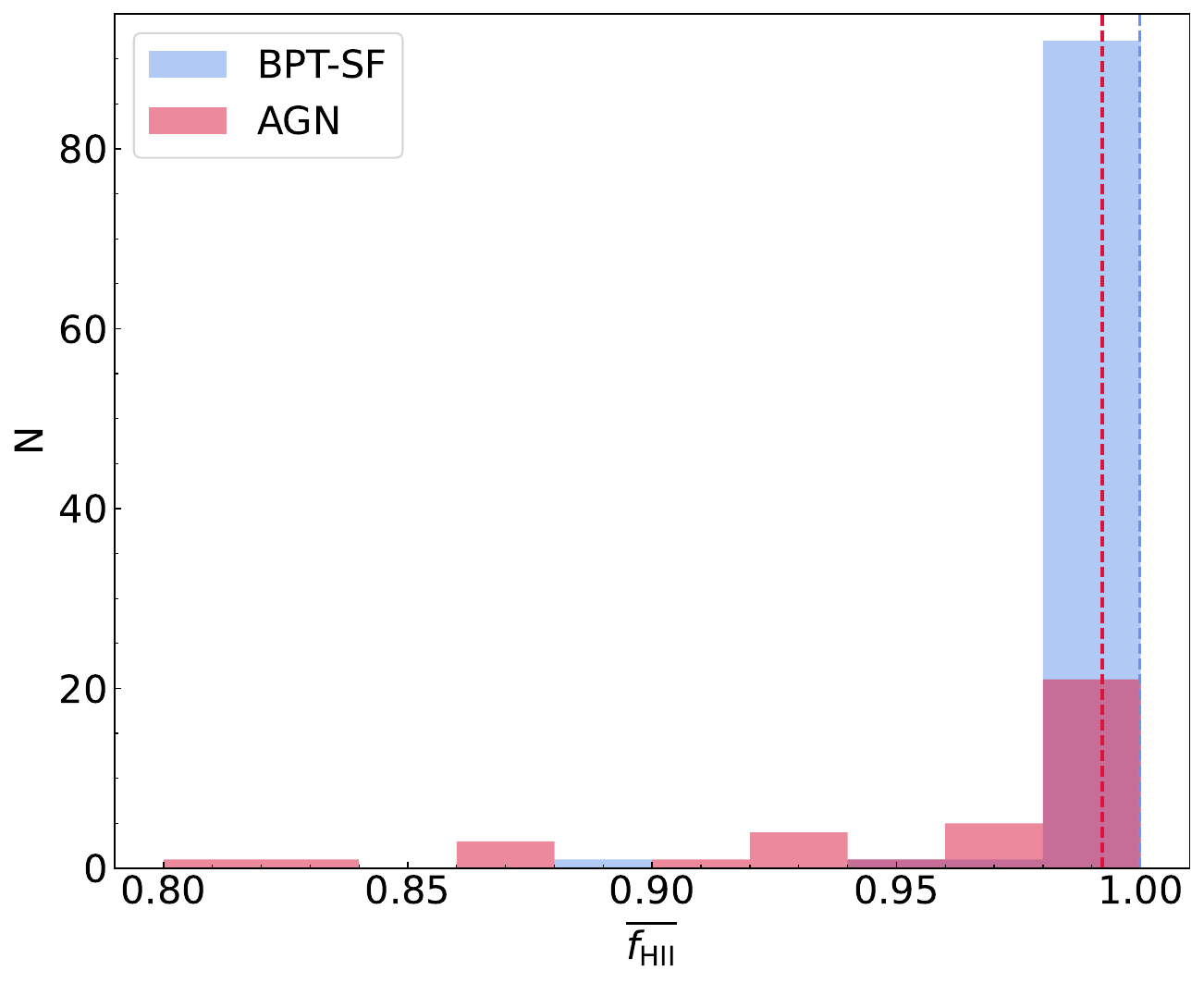}}
    \caption{The distribution of the H$\alpha$ luminosity-weighted $\hii$ region fraction $f_\hii$ calculated by \autoref{equ:mean_fhii} for BPT-SF galaxies (blue) and AGN-host galaxies (red). The blue (red) dashed line indicates the median $\overline{f_\hii}$ of BPT-SF (AGN-host) galaxies.}
    \label{fig:fhii_dis}
\end{figure}

\begin{figure}
{\includegraphics[width=\columnwidth]{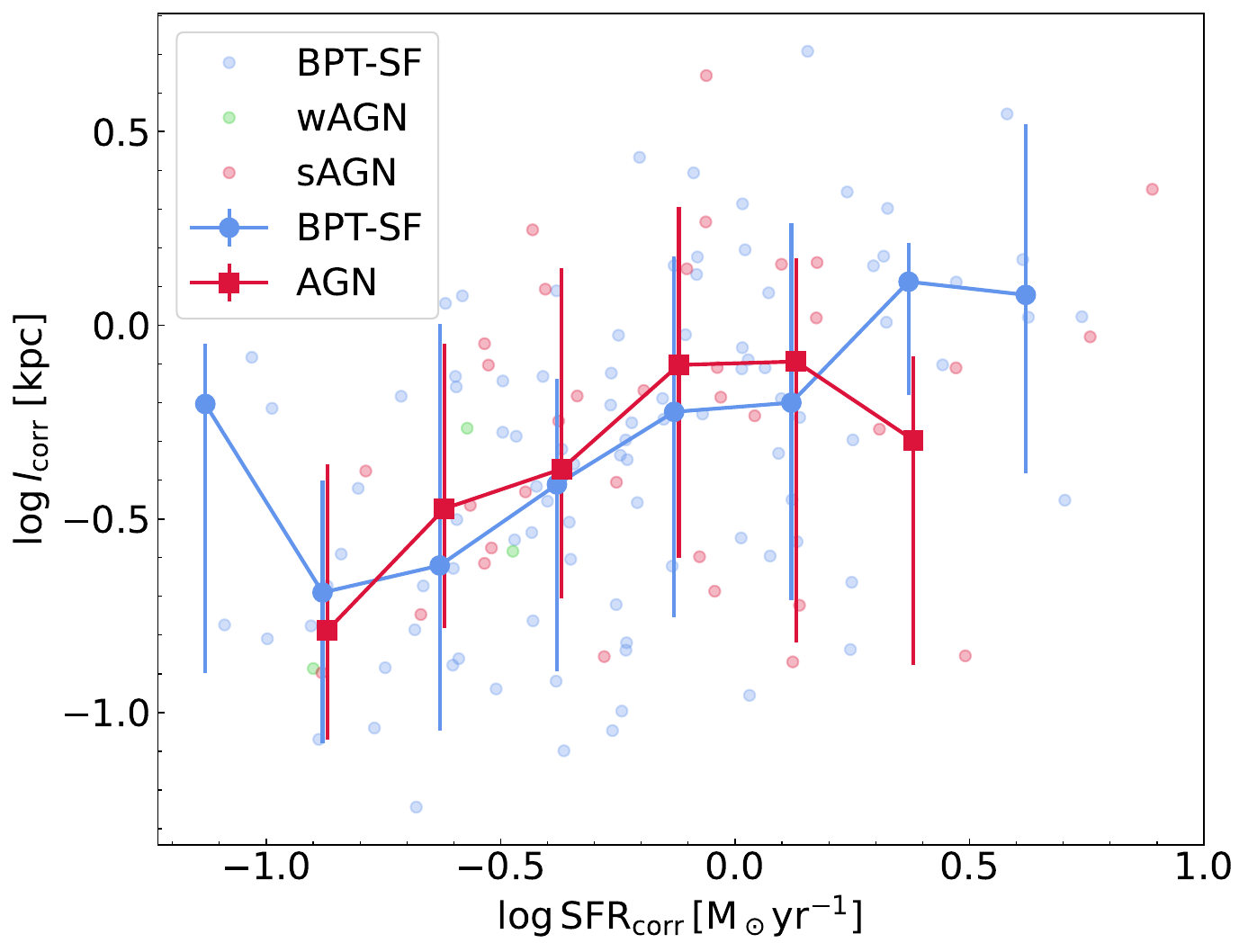}}
    \caption{The same as the right panel of \autoref{fig:l_corr_relation}, but showing the relation of $\lcor$ with the corrected SFR computed following the procedure described in \aref{app:sfr_corr}.}
    \label{fig:sfr_corr_l}
\end{figure}

As discussed in \autoref{sec:mass_sfr}, the SFR reported in the eDR3 catalog is directly converted from H$\alpha$ luminosity without any attempt to correct for AGN contributions. Accordingly, \citet{sanchez2023} point out that the SFR for AGN-host galaxies can only be treated as an upper limit. In this Appendix we quantify the level of AGN contamination following the procedure described in \citetalias{li2024}.

We firstly run \textsc{NebulaBayes} on our sample with the same spaxel masks and following the same procedure introduced in \autoref{sec:sample} and \autoref{sec:NB}. One output of this calculation for every spaxel is $f_\hii$, which describes the fraction of H$\alpha$ luminosity attributable to $\hii$ regions driven by stars, as opposed to AGN. We then calculate the H$\alpha$-luminosity weighted mean $f_\hii$:
\begin{equation}
\overline{f_\hii}=\frac{\sum_{i} L_{\mathrm{H}\alpha,i} f_{\hii,i}}{\sum_{i} L_{\mathrm{H}\alpha,i}},
\label{equ:mean_fhii}
\end{equation}
where $L_{\mathrm{H}\alpha,i}$ and $f_{\hii,i}$ are the dust-corrected H$\alpha$ luminosity and value of $f_\hii$ returned by \textsc{NebulaBayes} for the $i$th spaxel; the sum is over all non-masked spaxels. Thus $\overline{f_\hii}$ reflects our best estimate of the fraction of the total H$\alpha$ luminosity contributed by massive stars, and thus attributable to star formation. Values of $\overline{f_\hii}$ closer to unity indicate less contamination from AGN. \autoref{fig:fhii_dis} shows the distribution of $\overline{f_\hii}$ values for the BPT-SF (blue) and AGN-host (red) galaxies in our sample. The median value $\overline{f_\hii}$ for BPT-SF galaxies is close to unity, as expected, but even for AGN-host galaxies this falls only to 0.992. It is therefore clear that even for AGN-host galaxies, the AGN contribution to the overall H$\alpha$ luminosity is marginal. This is because most AGN-host galaxies in our sample are similar to the example shown in \autoref{fig:bpt_example}, where the AGN contributes significantly only in the centre, and is not the dominant ionising source throughout the galaxy.

To understand how correcting for AGN contamination might influence our final results, we re-estimate the SFR excluding the AGN contribution using our $f_\hii$ maps: 
\begin{equation}
\mathrm{SFR_{NB}}=\left[\sum_{i} L_{\mathrm{H}\alpha,i} f_{\hii,i}\right]\times \frac{7.9\times10^{-42}\,\mathrm{M}_\odot\mbox{ yr}^{-1}/(\mbox{erg s}^{-1})}{10^{0.25}}.
\label{equ:sfr_nb}
\end{equation}
Here  $L_{\mathrm{H}\alpha,i}$ and $f_{\hii,i}$ are the same as in \autoref{equ:mean_fhii}, and the factor of $10^{0.25}$ in the denominator is to correct from the \citet{salpeter1955} IMF to \citet{chabrier2003} IMF. It is important to note that, because many spaxels in our maps are masked due to low SNR, we cannot directly compare $\mathrm{SFR_{NB}}$ to the SFR derived from the catalogue, $\mathrm{SFR_{eDR3}}$ -- even in galaxies with no AGN contamination the latter is usually significantly larger because it is based on the integral H$\alpha$ luminosity, whereas $\mathrm{SFR_{NB}}$ includes only the H$\alpha$ flux arising from spaxels bright enough to pass our SNR cuts. Given this situation, our approach to estimating the AGN correction is to fit a linear relation between the $\log \mathrm{SFR_{NB}}$ and the $\log \mathrm{SFR_{eDR3}}$ for BPT-SF galaxies, where there is no AGN contamination, and use the same relation to calibrate $\mathrm{SFR_{NB}}$ to a corrected value $\mathrm{SFR_{corr}}$ for AGN-host galaxies. This amounts to assuming that we are missing the same fraction of the flux due to low SNR in both AGN-host and non-AGN galaxies, and if anything this assumption will tend to under-correct the AGN-host galaxies, since the generally smaller SFR in AGN hosts reduces the number of spaxels with SNR high enough to pass our cuts. We can therefore regard the value of $\mathrm{SFR_{corr}}$ we derive from this procedure as a lower limit on the SFR in AGN-host galaxies, with the uncorrected value $\mathrm{SFR_{eDR3}}$ representing an upper limit.

\autoref{fig:sfr_corr_l} shows the relation between the $\lcor$ and the corrected $\log \mathrm{SFR_{corr}}$; we generate this figure exactly as we generate the right panel of \autoref{fig:l_corr_relation}, but using $\mathrm{SFR_{corr}}$ instead of $\mathrm{SFR_{eDR3}}$. As in the right panel of \autoref{fig:l_corr_relation}, the BPT-SF and the AGN-host galaxies show similar trends. Using the same $t$-test procedure as described in \autoref{sec:results}, we find that the $p$-values with which we can reject the null hypothesis that the mean $\lcor$ of the AGN hosts equal to that of BPT-SF galaxies are 0.92, 0.40, 0.22, 0.41, 0.87, and 0.03 from the lowest SFR to the highest bin. Thus we obtain the same result as we did for $\mathrm{SFR_{eDR3}}$, which is that we cannot reject the null hypothesis. Given the small difference that AGN correction makes to our results, we elect to use the uncorrected SFR from the eDR3 catalogue in the main body of this work.


\bsp	
\label{lastpage}
\end{document}